\documentclass[numbers]{article}

\usepackage[final,dandb]{dermaconin}

\usepackage{longtable}
\usepackage{makecell}
\usepackage[utf8]{inputenc} % allow utf-8 input
\usepackage[T1]{fontenc}    % use 8-bit T1 fonts
\usepackage{hyperref}       % hyperlinks
\usepackage{url}            % simple URL typesetting
\usepackage{booktabs}       % professional-quality tables
\usepackage{pifont} % For tick and cross
\newcommand{\cmark}{\ding{51}}  % tick
\newcommand{\xmark}{\ding{55}}  % cross
\usepackage{amsfonts}       % blackboard math symbols
\usepackage{nicefrac}       % compact symbols for 1/2, etc.
\usepackage{xcolor}         % colors
\usepackage{wrapfig} % Put this in your preamble if not already included
\usepackage{arydshln}
\usepackage[font=small]{caption}  % or footnotesize, scriptsize, etc.
\usepackage{longtable}
\usepackage{amsmath}
\usepackage{multirow}
\usepackage{url}            % Proper formatting of URLs
\usepackage{doi}            % Clean DOI links
\usepackage{graphicx}  % In the preamble

\title{DermaCon-IN: A Multi-concept Annotated Dermatological Image Dataset of Indian Skin Disorders for Clinical AI Research}

\author{%
  Shanawaj S Madarkar\thanks{Equal contribution}\: \thanks{Department of Artificial Intelligence, Indian Institute of Technology Hyderabad, India}\: \thanks{Indian Navy}\footnotemark[2] \and
  \textbf{Mahajabeen Madarkar}\footnotemark[1]\:  \thanks{Department of Dermatology, S R Patil Medical College, India} \and
  \textbf{Madhumitha Venkatesh}\footnotemark[1]\: \footnotemark[2] \and
  Deepanshu Bansal\thanks{Department of Dermatology, S Nijalingappa Medical College, India} \and 
  Teli Prakash\footnotemark[5] \and
  Konda Reddy Mopuri\footnotemark[2] \and
  Vinaykumar MV\thanks{Department of Dermatology, Sri Chamundeshwari Medical College, Hospital \& Research, India} \and
  KVL Sathwika\thanks{Interns at Indian Institute of Technology Hyderabad, India} \and
  Adarsh Kasturi\footnotemark[7] \and
  Gandla Dilip Raj\footnotemark[7] \and
  PVN Supranitha\footnotemark[7] \and
  Harsh Udai\footnotemark[2] 
}

\begin{document}

\maketitle

\begin{abstract}
% AI in Dermatology is poised to transform clinical workflows by enabling scalable, image-based decision support across diverse populations and disease spectra. Realizing this potential, however, requires datasets that mirror the complexity of real-world dermatology, spanning tonal variation, diagnostic breadth, and clinical fidelity. Yet, most existing resources are constrained by narrow taxonomies, limited skin tone representation, and data pipelines decoupled from routine clinical practice. We introduce a prospectively curated dermatology dataset developed across outpatient clinics in South India, comprising over $5,450$ clinical images from $\sim3,000$ unique patients. Each image is annotated by board-certified dermatologists with over $240$ distinct diagnoses hierarchically structured under an etiologic Rook’s taxonomy aligned with epidemiologic data. Annotations include body site localization, 48 lesion-level clinical concepts describing morphology, texture, and pigmentation, diagnostic certainty, and Fitzpatrick and Monk skin tone scores. We benchmark the dataset with CNNs, Vision Transformers, and concept bottleneck models, demonstrating how anatomical context and concept-level reasoning drive more robust and equitable performance.
Artificial intelligence is poised to augment dermatological care by enabling scalable image-based diagnostics. Yet, the development of robust and equitable models remains hindered by datasets that fail to capture the clinical and demographic complexity of real-world practice. This complexity stems from region-specific disease distributions, wide variation in skin tones, and the underrepresentation of outpatient scenarios from non-Western populations. We introduce DermaCon-IN, a prospectively curated dermatology dataset comprising 5,450 clinical images from 3,002 patients across outpatient clinics in South India. Each image is annotated by board-certified dermatologists with 245 distinct diagnoses, structured under a hierarchical, aetiology-based taxonomy adapted from Rook’s classification. The dataset captures a wide spectrum of dermatologic conditions and tonal variation commonly seen in Indian outpatient care. We benchmark a range of architectures, including convolutional models (ResNet, DenseNet, EfficientNet), transformer-based models (ViT, MaxViT, Swin), and Concept Bottleneck Models to establish baseline performance and explore how anatomical and concept-level cues may be integrated. These results are intended to guide future efforts toward interpretable and clinically realistic models. DermaCon-IN provides a scalable and representative foundation for advancing dermatology AI.

\end{abstract}

\section{Introduction} \label{sec:introduction}

Skin diseases pose a significant global health challenge, affecting billions of individuals and ranking among the leading causes of disease burden. The Global Burden of Disease 2019~\cite{thakur2021burden_skin_india} study identified dermatological conditions as the \textit{fourth} leading cause of nonfatal morbidity worldwide. Common ailments such as fungal infections, acne, scabies, and eczema impact millions globally~\cite{karimkhani2017global}, underscoring the urgent need for improved diagnostic tools and equitable access to care.

Artificial intelligence (AI) has emerged as a promising solution for enhancing dermatological diagnosis and triage, particularly in resource-constrained regions with limited access to dermatologists. However, a critical bottleneck remains: the lack of representative training data that adequately captures diversity in disease presentations and skin tones based on regional relevance~\cite{daneshjou2022disparities}. Most AI models for dermatology to date have been developed and benchmarked using datasets predominantly sourced from North American, European, or Australasian populations~\cite{alipour2024skin,lopezperez2025generative}. This geographic skew has introduced performance biases, especially for underrepresented groups such as individuals, patients presenting with diseases more prevalent outside Western contexts, and those  with darker skin tones.

\begin{figure}[t]
  \centering
  \includegraphics[width=0.9\linewidth]{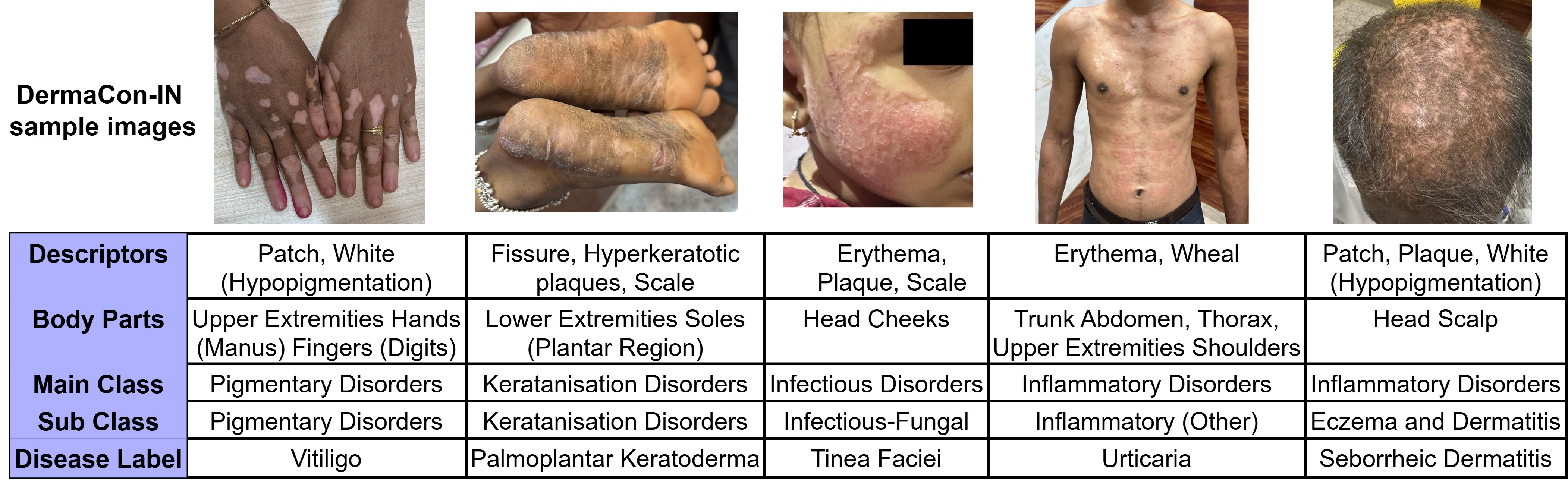}
  \captionsetup{font=scriptsize}
  \caption{Sample images from DermaCon-IN dataset with skin lesion descriptors, body parts, main class, sub class, and disease labels.}
  \vspace{-10pt}

  \label{fig:sample}
\end{figure}

 Recent work reveals the consequences of biased dermatology datasets~\cite{alipour2024skin,omalley2024representation}. Public benchmarks focus on pigmented lesions and melanoma, mirroring Western priorities, while overlooking common conditions in tropical regions, like fungal infections, scabies, pigmentary, and nutritional disorders~\cite{szeto2024dermatologic,thakur2021burden_skin_india,karimkhani2017global,hay2014global,johnson2020global}. A “one-size-for-all” approach fails across populations, demanding region-specific resources. Under-representation of darker skin tones compounds this gap:~\cite {daneshjou2022disparities} reports 30-40\% accuracy drop on darker skin in DDI, while trained on  Fitzpatrick17k's dataset (dominant lighter tones in the dataset). These biases lead to models that underperform on underrepresented phenotypes.
% Recent studies reveal the tangible consequences of this imbalance. Existing datasets exhibit a narrow disease focus. Publicly available benchmarks typically emphasize pigmented lesions and skin cancers, notably malignant melanoma, which reflect Western disease priorities. This leaves out high-prevalence conditions such as fungal infections, scabies, pigmentary disorders, and nutritional dermatoses that are dominant in tropical and developing regions. As highlighted by \citet{xie2023generalizability}, a ``one-size-fits-all'' dermatology dataset is unlikely to generalize across ethnicities, necessitating region-specific resources to support robust and equitable AI systems.
 
% Beyond disease disparity, For instance, \citet{daneshjou2022ddi} report a 30--40\% reduction in diagnostic accuracy for state-of-the-art skin lesion classifiers when evaluated on darker skin tones and rare conditions in the Diverse Dermatology Images (DDI) dataset. Likewise, the Fitzpatrick17k dataset---comprising 16,577 clinical images labeled by Fitzpatrick skin type---demonstrated a significant overrepresentation of lighter skin tones (Types I--III), echoing similar skews in widely used dermatology atlases. Such imbalances contribute to algorithmic bias, wherein models trained predominantly on lighter skin perform poorly on lesions in darker-skinned individuals.

To address these limitations, we introduce a new dermatology image dataset curated from Indian outpatient clinics. To the best of our knowledge, it is the first densely annotated dataset centered around Indian skin phototype and is designed to improve diversity in both disease coverage and skin tone representation for dermatological AI research. It complements existing datasets by capturing the phenotypic and pathological landscape of a population historically underserved in global medical AI efforts. In addition, the dataset also aims to support explainable modeling by reflecting how dermatologists diagnose, through the combined use of anatomical location and visual descriptors. The Key contributions of this work are as follows:
\vspace{-0.5em}

\begin{itemize}
    \item \textbf{South Asian Clinical and Phenotypic Representation.}  
    DermaCon-IN developed in South Asia reflects regional disease patterns, such as the high prevalence of infectious etiologies (fungal, viral, parasitic) observed in tropical outpatient settings~\cite{urban2021global,hay2020skin,shah2019himatnagar,balasubramanian2021andaman,jayanthi2017northchennai}. This contrasts with existing datasets dominated by inflammatory or neoplastic disorders common in Western contexts~\cite{Yakupu2023Burden,Huai2025Global}. The dataset also includes Fitzpatrick skin types IV--VI, which are typically underrepresented in existing resources~\cite{groh2021evaluating}, offering a path to reduce fairness gaps in clinically deployable AI models.

    \item \textbf{Multi-Concept Clinical Annotations.}  
    Each image is annotated with two independent sets of clinically meaningful metadata: precise anatomical locations and lesion-level descriptors that capture surface and morphological features of skin lesions (seen in Figure~\ref{fig:sample}). To the best of our knowledge, this is the first publicly available dataset to offer both annotation types at this scale and granularity, supporting structured supervision and interpretable modeling.

    \item \textbf{Clinically Aligned Hierarchical Labeling.} Disease labels are organized in a three-tier hierarchy: main diagnostic class, etiology-based subclass, and specific disease label. This structure is derived from Rook’s \textit{Textbook of Dermatology} (the clinical gold standard)~\cite{Griffiths2024Rook} and adapted to Indian dermatology practice. It mirrors diagnostic workflows in real-world settings, enabling both coarse- and fine-grained modeling.

    \item \textbf{Benchmarking for Classification and Interpretability.}  
    We provide baseline results for disease classification and for Concept Bottleneck Models (CBMs)~\cite{Koh2020Concept} that leverage the concept annotations. These benchmarks demonstrate the dataset’s relevance for both predictive accuracy and to evaluate whether models are learning medically meaningful concepts in alignment with expert reasoning.
\end{itemize}

\section{Related Work}\label{sec:Related Work}
\vspace{-9pt}
\begin{table}[h]
\vspace{-9pt}
\scriptsize
\centering
\captionsetup{font=scriptsize}
\caption{Comparative Survey of existing Dermatology Datasets available for AI research [Columns: \textbf{A}: Neoplasm \& Tumors Centric, \textbf{B}: Broad Skin Disease Spectrum, \textbf{C}: Dermoscopic Single Lesion Focus, \textbf{D}: Real-time Multi-lesion Multi-Focus, \textbf{E}: Body Part, \textbf{F}: Lesion Descriptor, \textbf{G}: Rook's classification labels]}
\setlength{\tabcolsep}{3pt} 
\begin{tabular}{l|cc|cc|ccc|cc|c|cc}
\toprule
\textbf{Dataset } &
\multicolumn{2}{c|}{\textbf{\makecell{Disease\\ Distribut.}}} &
\multicolumn{2}{c|}{\textbf{\makecell{Acquisit.\\ Type}}} &
\multicolumn{3}{c|}{\textbf{\makecell{Dense\\ Annotat.}}} &
\multicolumn{2}{c|}{\textbf{Source of Images}} &
\textbf{\makecell{Skin\\Tone}} &
\multicolumn{2}{c}{\textbf{Classes}} \\
 & A & B & C & D & E & F & G & \makecell{Web Scraped\\(Atlas)} & \makecell{Geographic\\Location} & Present & \#Images & \makecell{Hierarchical\\\#level[\#count]} \\
\midrule
ISIC Archive~\cite{isic2020} & \cmark & \xmark & \cmark & \xmark & \cmark & \xmark & \xmark & \xmark & Europe & \xmark & $\sim$485,000 & 1[9] \\
HAM10000~\cite{tschandl2018ham10000} & \cmark & \xmark & \cmark & \xmark & \cmark & \xmark & \xmark & \xmark &Austria,Australia & \xmark & 10,015 & 1 [7] \\
DERM12345~\cite{yilmaz2024derm12345} & \cmark & \xmark & \cmark & \xmark & \xmark & \xmark & \xmark & \xmark & Türkiye & \xmark & 12,345 & 3 [5,15,38] \\
BCN20000~\cite{hernandez2024bcn20000} & \cmark & \xmark & \cmark & \xmark & \cmark & \xmark & \xmark & \xmark & Spain & \xmark & 18,946 & 1 [8] \\
PH2~\cite{mendonca2013ph2} & \cmark & \xmark & \cmark & \xmark & \xmark & \cmark & \xmark & \xmark & Portugal & \xmark & 200 & 1 [3] \\
PAD-UFES-20~\cite{pacheco2020pad} & \cmark & \xmark & \xmark & \cmark & \cmark & \xmark & \xmark & \xmark & Brazil & \xmark & 1,612 & 1 [6] \\
DDI~\cite{daneshjou2022disparities} & \cmark & \xmark & \xmark & \xmark & \cmark & \xmark & \xmark & \xmark & USA & \cmark & 656 & 1[78] \\
Derm7pt~\cite{kawahara2018sevenpoint} & \cmark & \xmark & \cmark & \xmark & \xmark & \cmark & \xmark & \cmark & Italy & \xmark & 1,011 & 2 [5, 20] \\
\midrule
Fitzpatrick17k~\cite{groh2021evaluating} & \xmark & \cmark & \xmark & \cmark & \xmark & \xmark & \xmark & \cmark & -- & \cmark & 16,577 & 3 [3,9,114] \\
SD-198~\cite{sun2016benchmark} & \xmark & \cmark & \xmark & \cmark & \xmark & \xmark & \xmark & \cmark & -- & \xmark & 6,584 & 1[198] \\
SkinCon~\cite{daneshjou2022skincon} & \xmark & \cmark & \xmark & \cmark & \xmark & \cmark & \xmark & \cmark & -- & \cmark & 3,886 & \xmark \\ 
PASSION~\cite{gottfrois2024passion} & \xmark & \cmark & \xmark & \cmark & \cmark & \xmark & \xmark & \xmark & Africa & \cmark & 4,901 & 1[4] \\
SCIN~\cite{ward2024creating} & \xmark & \cmark & \xmark & \cmark & \cmark & \xmark & \xmark & crowd-sourced & USA & \cmark & 10,000+ & 1 [419] \\
\midrule
DermaCon-IN & \xmark & \cmark & \xmark & \cmark & \cmark & \cmark & \cmark & \xmark & South India & \cmark & 5,450 & 3 [8,19,254] \\
\bottomrule
\end{tabular}
\label{tab:lit}
\vspace{-10pt}
\end{table}

\paragraph{Neoplasm-Centric Benchmarks:}Early dermatology AI models were trained on datasets focused on neoplasms and tumours, such as the ISIC~\cite {isic2020}, HAM10000~\cite{tschandl2018ham10000}, DERM12345~\cite{yilmaz2024derm12345}, etc, as discussed in Table~\ref{tab:lit}. These primarily contain dermoscopic images and omit common diseases like infectious and inflammatory disorders. Dermoscopic imaging focusing on a single lesion further abstracts clinical variability in lighting, context, and lesion complexity, limiting real-time applicability.

\vspace{-3pt}
\textbf{Atlas-Sourced Clinical Datasets:} SD-198~\cite{sun2016benchmark} and Fitzpatrick17k~\cite{groh2021evaluating} introduced clinical (non-dermoscopic) photographs to broaden the coverage of disease spectrum. However, both are derived from educational atlases (like DermNet), not clinical repositories, yielding limited annotations. Moreover, the Fitzpatrick17k ~\cite{groh2021evaluating} dataset excludes several prominent diseases, including Fungal and Viral infections, and has skewed tonal variation of over 75\% belonging to Types I–III. 
\vspace{-10pt}
\paragraph{Fairness-Focused Collections:}Datasets like DDI ~\cite{daneshjou2022disparities} and PASSION ~\cite{gottfrois2024passion} emphasise tonal representation but trade off diagnostic breadth. DDI includes fewer than 80 disease labels, of which neoplastic or pigmentary offer a larger contribution. PASSION~\cite{paszke2019pytorch} has pediatric participants' images across only four conditions (eczema, fungal infections, scabies, impetigo), selected for regional prevalence. SCIN~\cite{ward2024creating}dataset, on the other hand, introduces crowd-sourced images, expanding coverage to common non-neoplastic conditions but mirrors U.S. disease patterns~\cite{laughter2020burden}, thus under-representing both high-burden infectious, pigmentary, etc, disorders seen in global contexts and darker skin tones. 
\vspace{-10pt}
\paragraph{Our Dataset in Context:}\label{sec:datsetcontext}
\begin{wrapfigure}{r}{0.45\linewidth}
% \scriptsize
  \centering
  \vspace{-10pt}
  \includegraphics[width=\linewidth]{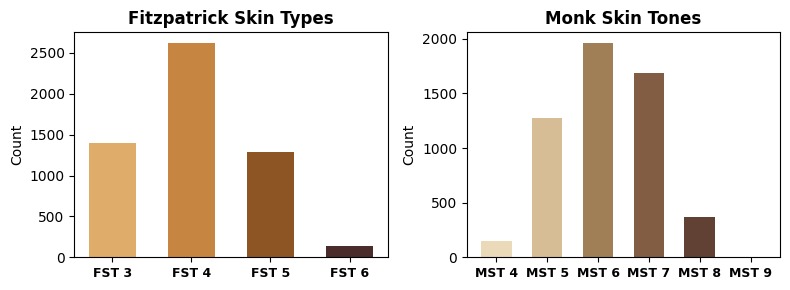}
  \vspace{-10pt}
  \captionsetup{font=scriptsize}
  \caption{FST and MST distribution of skin tones of subjects in DermaCon-IN dataset}
  \vspace{-15pt}
  \label{fig:FSTMST}
\end{wrapfigure}
In South Asia, outpatient dermatology is dominated by inflammatory, infectious, pigmentary, and appendageal diseases~\cite{urban2021global}, yet remains underrepresented in existing datasets. We address this gap with a dataset of $5{,}450$ high-resolution clinical images collected prospectively from 3{,}002 Indian patients. It covers the disease spectrum aligned with Indian and global burden data. Each image retains anatomical context and is annotated by board-certified dermatologists with standardised diagnosis, as well as Fitzpatrick and MST skin tone ratings, which align with patterns observed in the Indian context~\cite{sachdeva2009fitzpatrick, sarkar2019randomised,hourblin2014skin}. The distribution of which is shown in Figure ~\ref{fig:FSTMST}. Unlike prior work such as SkinCon~\cite{daneshjou2022skincon}, which retrofitted a set of lesion descriptors onto existing datasets, we capture both lesion and anatomical concepts at source and leverage the full concept set for statistical validation and model benchmarking.

\section{Data Collection Methodology}
% \vspace{-8pt}
\subsection{Clinical Setting and Data Sources}
\vspace{-5pt}
The DermaCon-IN dataset was developed via multi-institutional collaboration involving three tertiary-care hospitals and affiliated regional clinics across North Karnataka, South India. The cohort represents a demographically and geographically diverse outpatient population from Karnataka, Maharashtra, Goa, and Andhra Pradesh. Data were collected between 2024 and 2025 under institutionally approved ethical protocols.

\vspace{-8pt}
\subsection{Image Acquisition Protocol}
\vspace{-5pt}
Image capture was designed to mirror real-world dermatology workflows, emphasizing both lesion detail and the broader anatomical region. Photographs were taken using high-resolution smartphone devices, viz, 108MP Android, 48MP iPhone Pro, and 12–36MP cameras, under ambient clinical lighting. Images include the affected body part with surrounding skin to preserve anatomical context and support spatial modeling. A standardized protocol guided acquisition across sites, allowing relevant variations in angle, distance, and lighting to reflect clinical realism, unlike prior datasets focused on dermoscopic or tightly cropped views.

\vspace{-8pt}
\subsection{Inclusion and Exclusion Criteria}
\vspace{-5pt}
Patients of all ages with clinically confirmed dermatologic conditions were included (with consent), contingent on diagnostic agreement by two board-certified dermatologists or follow-up validation. Only images meeting gradability standards and accompanied by complete metadata, including diagnosis, anatomical region, Fitzpatrick and Monk tone ratings, demographics, and diagnostic confidence, were retained. Exclusion criteria included poor image quality, visual obstructions (e.g., tattoos, accessories), metadata gaps, or ambiguous diagnoses, and patients unwilling to participate.

\vspace{-8pt}
\subsection{Annotation Process}
\vspace{-5pt}
The entire dataset was annotated by four board-certified Dermatologists with clinical experience of 11 years, 3 years, 3 years, and 1 year, respectively. The entire dataset was divided into four smaller subsets for labelling by these doctors based on the availability of the dermatologist. Labels followed a three-level disease taxonomy informed by Rook’s Classification~\cite{Griffiths2024Rook}, which is considered a gold standard in Dermatology (Refer to Supplementary Sec. A). Annotations also included $47$ lesion-level descriptors and $49$ body part locations, along with patient metadata not linked to patients' privacy. Discrepancies were resolved via consensus or adjudication by a third expert.

\begin{wrapfigure}{r}{0.25\linewidth}
% \scriptsize
  \centering
  \vspace{-35pt}
  \includegraphics[width=\linewidth]{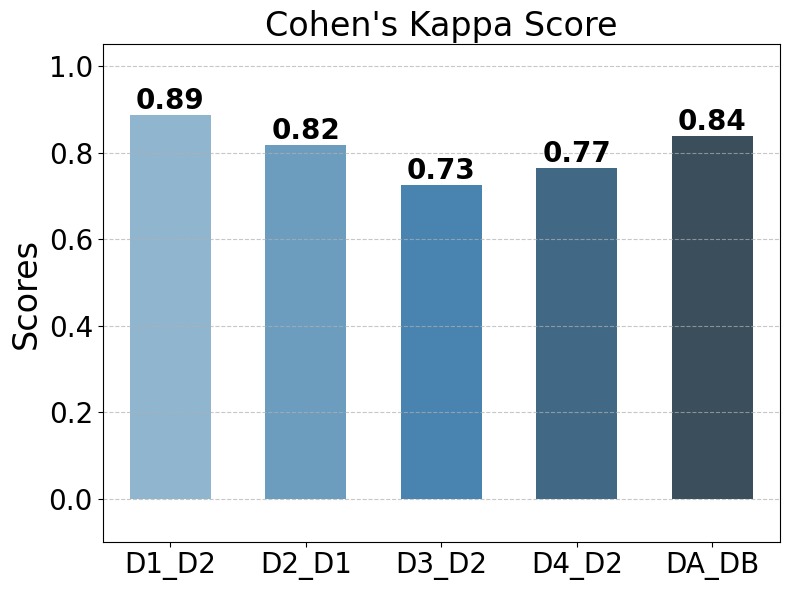}
  \vspace{-10pt}
  \captionsetup{font=scriptsize}
  \caption{Cohen's Kappa scores for cross validation of annotations provided by $4$ doctors (D1,D2,D3,D4)}
  \vspace{-15pt}
  \label{fig:kappa}
\end{wrapfigure}

\vspace{-8pt}
\subsection{Quality Control and Inter-Rater Agreement}
\vspace{-5pt}
To ensure consistency, 10–15\% of each annotator’s batch was randomly reviewed by another dermatologist. Inter-rater reliability, measured using Cohen’s Kappa, achieved a score of $0.84$ (Figure~\ref{fig:kappa}), aligning with accepted clinical annotation standards. Skin tone ratings were independently assigned by the set of trained experts, which was verified for consensus by the dermatologist. Anatomical site labels were validated using structured region maps.

\vspace{-8pt}
\subsection{Final Dataset Composition}
\vspace{-5pt}
The final composition contains $5{,}450$ high-resolution JPEG images across $8$ top-level etiologic classes,  $19$ clinically meaningful subclasses, and $245$ fine-grained disease labels. Each sample includes dense metadata: hierarchical disease labels, body parts, skin lesion descriptors, Fitzpatrick~\cite{fitzpatrick1975soleil} and Monk skin tone~\cite{schumann2023consensus} scores, diagnostic certainty, and image gradability. We consider $49$ body parts and 
$47$ lesion descriptors as concepts, which account for $96$ unique concepts.

\section{Dataset Overview and Statistics}\label{sec:datset_overview}
\vspace{-0.5em}

\begin{wrapfigure}{r}{0.5\textwidth}
  \centering
  \vspace{-25pt}
\includegraphics[width=0.5\textwidth]{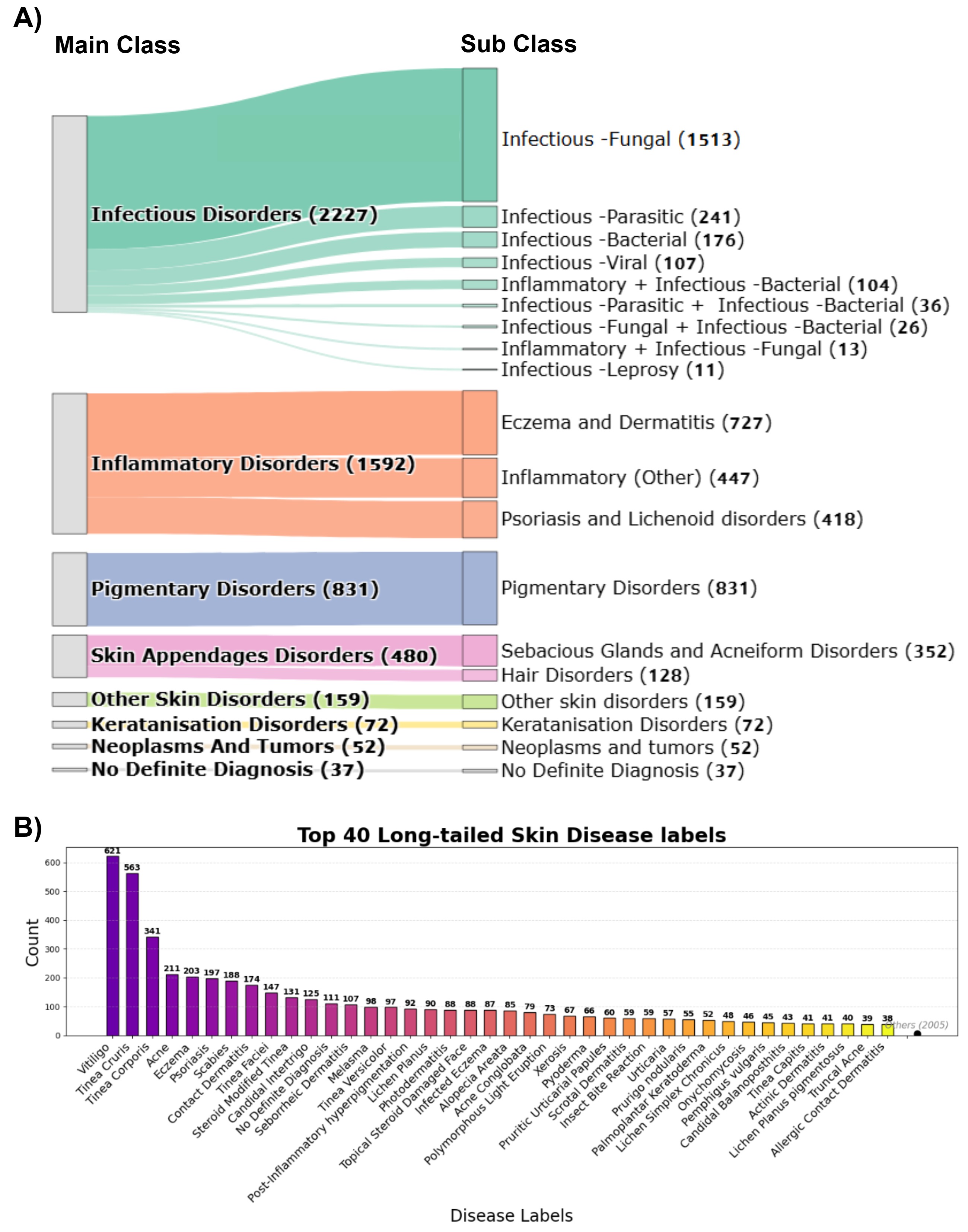}
  \vspace{-14pt}
  \captionsetup{font=scriptsize}
  \caption{(A) Hierarchical structure of dermatological annotations (according to Rook's classification) showing main classes and their corresponding sub-classes with image counts (in bracket). (B) Distribution of the top 40 most frequent disease labels, illustrating the long-tailed nature of the dataset.}
    \vspace{-15pt}
  \label{fig:Hierarchical Disease distribution}
\end{wrapfigure}

\paragraph{Diagnostic structure and class granularity:} 
DermaCon-IN is organized around an aetiologically informed, clinically validated taxonomy rooted in Rook’s~\cite {Griffiths2024Rook} classification and aligned with ICD-11~\cite{who2019icd11}. The dataset includes $8$ high-level diagnostic categories ranging from Infectious to Neoplastic, including No Definite Diagnosis, reflecting the full spectrum of dermatological conditions prevalent in South Asian outpatient settings~\cite{urban2021global}. These main classes are further expanded into 19 subclasses to capture hybrid and co-occurring disease states. For example, subclass combinations such as Inflammatory + Infectious (Bacterial) reflect polymorphic real-world presentations of superimposed two diseases. Each top-level category is populated with sufficient training instances for stratified benchmarking; notably, high-burden groups like fungal infections and eczema dominate in volume, while rare but clinically significant categories, e.g., keratinisation disorders remain represented with enough density to enable few-shot generalisation. The Disease label reflects $245$ distinct disease types, which follow a long-tailed distribution with a log-normal fit to leaf-node frequencies, yielding an exponent of $1.8$, as expected in real-world clinical settings. To better understand label diversity in the dataset, refer Figure~\ref{fig:Hierarchical Disease distribution}. This visualization underscores the need for imbalance-aware learning strategies.

\vspace{-10pt}
\paragraph{Lesion Descriptor annotations:}   
The visual descriptors of skin lesions 
\begin{wrapfigure}{r}{0.6\textwidth}
  \centering
  \vspace{-20pt}
  \includegraphics[width=0.6\textwidth]{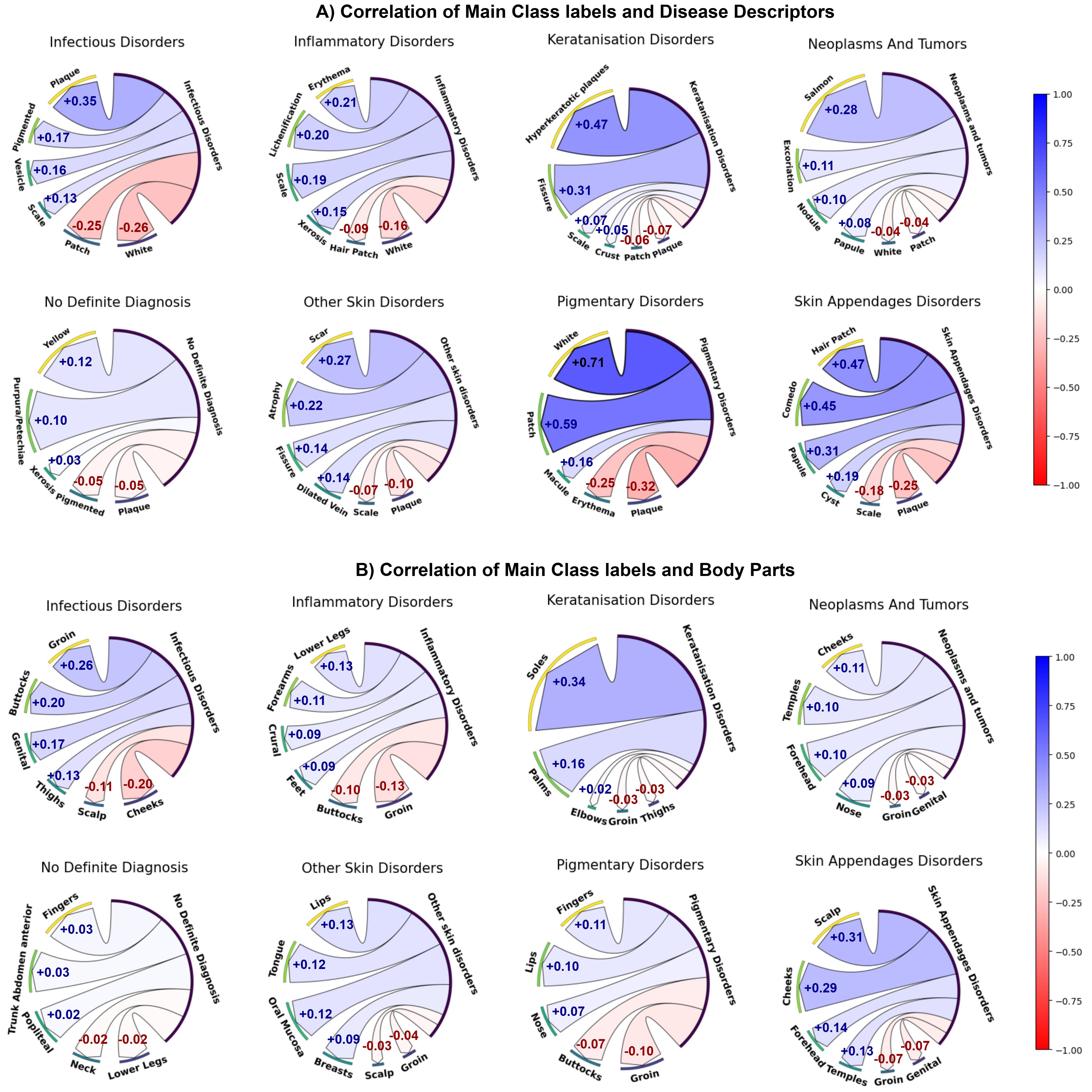}
  \captionsetup{font=scriptsize}
  \caption{Chord diagrams of Pearson correlations (\(r\)) between eight disease classes and two concept types: \textbf{(A)} lesion descriptors and \textbf{(B)} body regions. Ribbon width represents within-class correlation strength, and colour encodes direction and magnitude (dark blue = positive; red = negative). Numbers on ribbons indicate \(r\). Top-\(k\) positive and negative associations per class highlight distinctive clinical and anatomical patterns across disease groups.}
  % Chord diagrams depicting the most informative concept-class correlations for each main skin disease class. (A) Correlations between disease classes and clinical descriptors (e.g., morphology, texture, pigmentation); (B) correlations between disease classes and annotated body regions. For both, Pearson correlation coefficients were computed, and the top-$k$ positive and bottom-$k$ negative associations per class were selected. Ribbon widths represent correlation magnitude, and color encodes direction and strength (blue: positive, red: negative), highlighting characteristic clinical and anatomical patterns across disease groups.
  \vspace{-40pt}
  \label{fig:ChordDiagrams}
\end{wrapfigure}
annotated in DermaCon-IN follow an established dermatological vocabulary commonly used by clinicians to describe skin lesions in line with extant clinical literature~\cite{bolognia2017dermatology}. Inspired by frameworks~\cite{daneshjou2022skincon} previously validated in dermatology AI literature, we consulted board-certified dermatologists to select clinically meaningful descriptors aligned with disease patterns frequently encountered in Indian outpatient settings. Rather than assigning fixed concepts per disease category, each image was independently annotated, reflecting the realistic variability in lesion appearance across patients, body regions, and skin types. Statistical analysis in Figure~\ref{fig:ChordDiagrams} of concept-disease correlations validated this flexible annotation strategy that supports concept-supervised learning approaches. Details of Descriptors utilized for annotation are provided in the Datasheet (Supplementary Material).

\vspace{-10pt}
\paragraph{Anatomical Site annotations:} We retain full views of affected anatomical regions wherever feasible. Partial cropping is applied in cases when identifiable facial features are present, in line with ethical guidelines to ensure patient privacy while preserving anatomical context reflecting how skin conditions are typically encountered in practice. This anatomical framing allows models to learn spatial priors: for instance, tinea capitis typically affects the scalp, while palmoplantar keratoderma appears on the palms and soles. 

Such distributional cues are vital for accurate real-world diagnosis but are often lost in lesion-only images. Our dataset includes annotations for $49$ distinct body sites, organised using both precise clinical terms and broader regional tags (e.g., trunk, extremities, scalp). A List of body parts annotated is provided in the the Datasheet (Supplementary Material).

\vspace{-7pt}
\paragraph{Need for integration of Anatomical Context and Lesion Descriptors:}~\label{needforbothconcepts}
 Clinically, the same lesion type can indicate different diagnoses depending on its location; for example, a scaly plaque on the scalp might point to psoriasis, whereas a scaly plaque on the foot can suggest a fungal infection (tinea pedis). By jointly annotating lesion features (e.g., scales, plaques) and their anatomical context (e.g., scalp, foot), our dataset enables models to better reflect dermatologists' real-world diagnostic approach.
 
\vspace{-1pt}
\textbf{Population structure and label density:} The dataset reflects age and sex demographics typical of South Indian outpatient settings. The Dataset also adopts both Monk Skin Tone (MST)~\cite{schumann2023consensus} and Fitzpatrick~\cite{fitzpatrick1975soleil} scales, addressing the limitations of Fitzpatrick’s UV-response bias. MST offers a perceptual alternative capturing wider tonal representation, especially for darker skin types, as the tonal scale has increased variation. The combined annotation aligns with Indian phenotypic distributions (MST 4–9, Fitzpatrick 3–4).

\vspace{-5pt}
\paragraph{Statistical validation of concept and Anatomical coherence with Disease Labels:} We examined Pearson correlation coefficients between (a) Disease descriptors and disease categories, and (b) anatomical body regions and disease categories, (refer Figure~\ref{fig:ChordDiagrams}) to assess the biological plausibility and interpretability of our clinical annotations. Each chord diagram visualizes these relationships, where ribbon \textbf{width} indicates the strength of association between a concept and the class (within-class correlation) and ribbon \textbf{color} (dark blue~$\rightarrow$~red) encodes the strength of correlation across classes (dark blue = strong positive; red = negative). Numeric labels on ribbons denote the actual correlation coefficients.

For instance, under \emph{Pigmentary Disorders}, the descriptor \emph{White} shows a moderately wide ribbon and dark-blue color (\(r = +0.71\)), reflecting a strong and distinctive association both within the class and relative to other classes. In contrast, \emph{Hyperkeratotic plaques} under \emph{Keratinization Disorders} display a wider but lighter-blue ribbon (\(r = +0.47\)), suggesting it is more class-specific but less distinctive across classes.
Overall, the chord diagrams reveal statistically meaningful associations that align well with established dermatological knowledge. For instance,  positive (high) correlations are observed between \textit{erythema}, \textit{vesicle}, and \textit{scale} with inflammatory disorders, and between \textit{hyperkeratotic plaques} and keratinisation disorders. 
Similarly, \textit{white patches} and \textit{pigmented lesions} show high positive associations with pigmentary and infectious disorders, respectively, underscoring the dermatologic specificity of the disease descriptors in general. Anatomical correlations further reinforce clinical fidelity. Keratinisation disorders predominantly localize to the \textit{soles} and \textit{palms}, consistent with plantar keratoderma patterns. Skin appendageal disorders, such as acne and seborrheic dermatitis, are strongly associated with sebaceous-rich zones like the \textit{scalp} and \textit{cheeks}~\cite{bolognia2017dermatology,rapini2007dermatology}. These findings validate the anatomical tropisms encoded in the dataset.

\section{Challenges, Opportunities \& Limitations}\label{sec:Challenges_Opportunities_ Limitations} The dataset presents a range of challenges and opportunities that stem from the inherent complexity of real-world clinical data, offering practical constraints and avenues for robust model development:

\vspace{-10pt}
\paragraph{Resolution heterogeneity:} 
\begin{wrapfigure}{r}{0.38\linewidth}
% \scriptsize
  \centering
  \vspace{-15pt}
  \includegraphics[width=\linewidth]{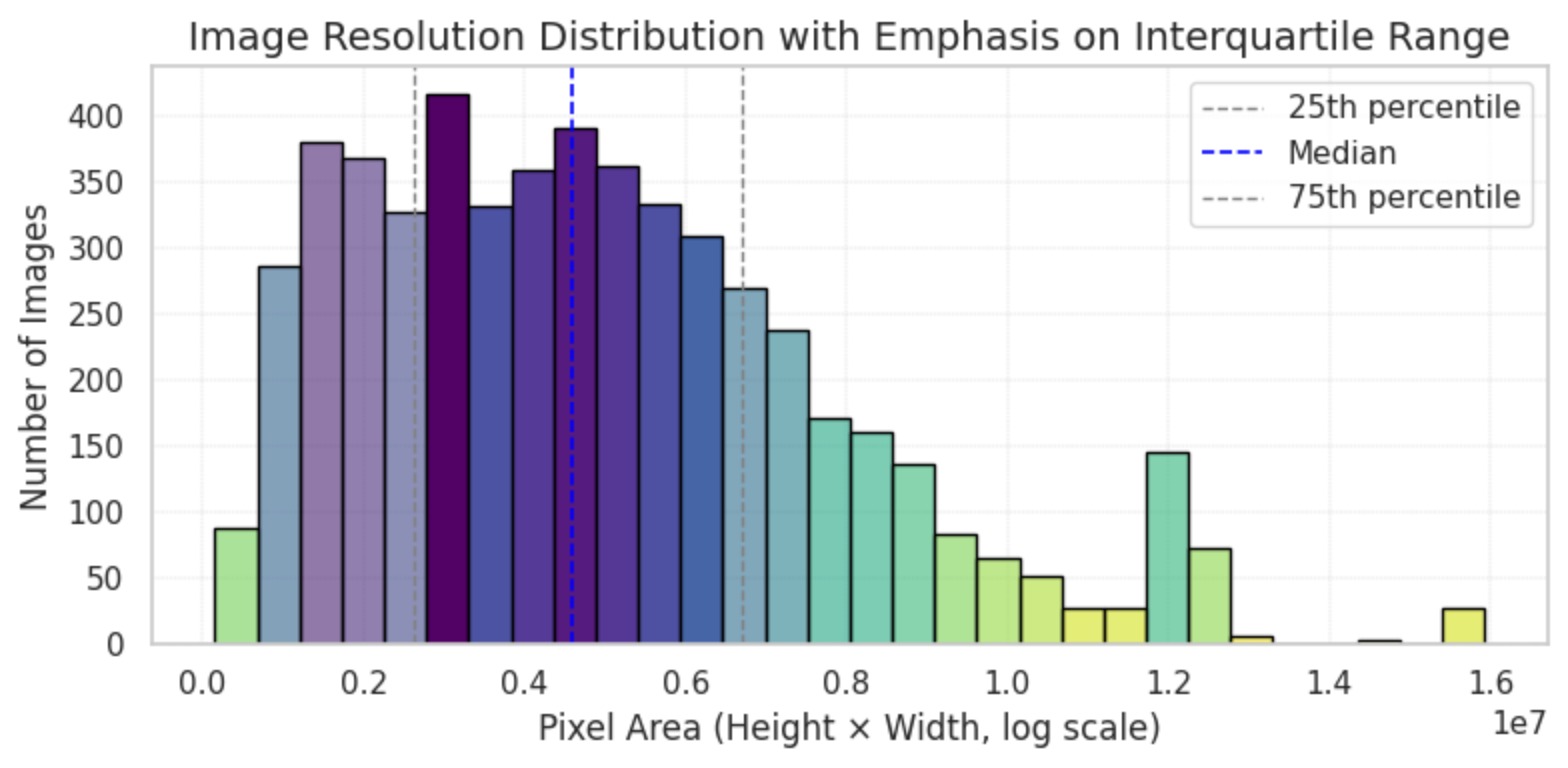}
  \vspace{-10pt}
  \captionsetup{font=scriptsize}
  \caption{Resolution distribution of images in DermaCon-IN dataset}
  \vspace{-10pt}
  \label{fig:res}
\end{wrapfigure}
Images were cropped post-acquisition to remove garments and background clutter where feasible, though incidental artifacts (e.g., jewelry) remain in some cases. The resulting variability in resolution, arising from diverse capture devices and aspect ratios, is a natural outcome, but advantageous. It reflects real-world conditions, where patients may crop or capture images themselves, and encourages model robustness to such variations. Image heights range from $296$ to $4{,}608$ pixels and widths from $346$ to $4{,}608$ pixels, with a mean resolution of $2{,}300$x$2{,}057$ pixels. The average image area is $4.97$M pixels ($\pm$3.01M), with an interquartile range of $2.62$M–$6.69$M pixels (Figure~\ref{fig:res}), indicating consistently high-fidelity input for fine-grained modeling.
\vspace{-5pt}
\paragraph{Hierarchical Labels and Clinical Distribution:}
\begin{itemize}
\vspace{-5pt}
    \item \textbf{Class Imbalance:} Our hierarchical labelling across main and sub-classes reflects true outpatient frequencies, common conditions are well-represented, while others occur proportionally less. This mirrors real-world practice and enables models to learn from naturally occurring clinical distributions.
    \vspace{-3pt}
    \item \textbf{Long-tailed Disease Labels:} The dataset embraces the inherent long-tailed nature of dermatological diagnoses, where few diseases are prevalent and many are rare. This structure presents a valuable opportunity to train models that generalise across the full clinical spectrum, including rare disease categories.
    \vspace{-3pt}
   \item \textbf{Multi-disease co-occurrences:} A subset shows concurrent lesions from multiple disease types on the same anatomical site (e.g., \textit{Inflammatory+Fungal}, \textit{Fungal+Bacterial}). We represent these as dedicated subclasses to help models disentangle overlapping pathologies for the main-class label we follow dermatologists’ treatment-priority logic, e.g., an infected eczema is assigned to the Infectious main class (not Inflammatory) because initial management targets the infection. Such cases, though less frequent, are observed in clinical settings, and our targeted misclassification analysis for these samples is provided in the Supplementary Sec. \textbf{B2}.

\end{itemize}
\vspace{-10pt}
\paragraph{Instance-level concepts:} These concepts describe what is visually observed in each image, such as scaling and erythema, and introduce two key characteristics that make DermaCon-IN a rich dataset for advancing clinical AI:
\vspace{-5pt}
\begin{itemize}
    \item \textbf{Class-agnostic semantics:} Descriptors (used as concepts) are shared across classes and are not rigidly tied to any single diagnostic category (e.g., \textit{scaling} alone $\rightsquigarrow$ ichthyosis, whereas \textit{scaling + erythema} $\rightsquigarrow$ psoriasis), with diagnostic meaning arising from their combinations. This invites the development of models capable of compositional and context-aware reasoning.
    \vspace{-3pt}
    \item \textbf{Long-tailed distribution:} The natural skew in concept frequencies mirrors real-world prevalence, where rare but critical findings coexist with common patterns. This creates opportunities to tackle challenges in multi-concept learning and rare concept detection, core problems in clinical AI.
\end{itemize}
\paragraph{Limitations.}\label{sec:limitations} In accordance with ethical considerations, all images were anonymized by masking identifiable facial features, such as the eyes, and cropping facial regions where necessary. While essential for protecting patient identity, this may limit the model’s ability to accurately learn or detect diseases that primarily manifest on the face.

\section{Benchmarking with Models}\label{sec:benchmarking}
\begin{table}[t]
\scriptsize
\centering
\captionsetup{font=scriptsize}
\caption{Performance benchmarking of the proposed dataset on the $ 8$-main class diagnosis classification task, conducted using both CNN- and ViT-based standard architectures. All metrics are averaged over $5$ random seeds.}
\label{tab:model_benchmark}
\setlength{\tabcolsep}{4pt}  % Default is 6pt; reduce as needed
\begin{tabular}{l c c c c c c c}
\toprule
\textbf{Model} & \textbf{Pre-trained}  & \textbf{Accuracy} & \textbf{Balanced Acc.} & \textbf{Precision} & \textbf{Sensitivity} & \textbf{F1 Score} \\
\midrule
ResNet50~\cite{he2016deep} & - &  $47.45_{\pm0.40}$ & $23.93_{\pm0.01}$ & $46.59_{\pm0.51}$ & $47.44_{\pm0.41}$ & $46.43_{\pm0.60}$ \\
DenseNet121~\cite{huang2017densely} & - &  $49.30_{\pm0.30}$ & $25.31_{\pm0.01}$ & $47.49_{\pm0.86}$ & $49.30_{\pm0.29}$ & $48.17_{\pm0.74}$ \\
ResNet50~\cite{he2016deep}  & ImageNet &  $64.31_{\pm0.22}$ & $38.77_{\pm1.40}$ & $63.41_{\pm0.29}$ & $64.31_{\pm0.23}$ & $63.31_{\pm0.25}$ \\
DenseNet121~\cite{huang2017densely} & ImageNet & $65.20_{\pm0.48}$ & $37.31_{\pm0.01}$ & $64.62_{\pm0.45}$ & $65.20_{\pm0.48}$ & $64.37_{\pm0.35}$ \\
EffNet-B4~\cite{tan2020efficientnet} & ImageNet & $64.28_{\pm0.34}$ & $35.58_{\pm0.01}$ & $63.53_{\pm0.64}$ & $64.27_{\pm0.34}$ & $63.38_{\pm0.39}$ \\
\midrule
ViT-B/16-224~\cite{dosovitskiy2020image} & ImageNet & $64.09_{\pm1.03}$ & $34.56_{\pm0.01}$ & $62.59_{\pm1.03}$ & $62.88_{\pm1.67}$ & $62.98_{\pm1.02}$ \\
ViT-B/16-384~\cite{dosovitskiy2020image} & ImageNet & $66.95_{\pm0.19}$ & $35.78_{\pm0.02}$ & $65.39_{\pm0.13}$ & $66.95_{\pm0.20}$ & $65.78_{\pm0.06}$ \\
MaxViT-B/512~\cite{tu2022maxvit} & ImageNet & $66.92_{\pm0.48}$ & $36.07_{\pm0.01}$ & $66.30_{\pm0.80}$ & $66.92_{\pm0.48}$ & $65.90_{\pm0.73}$ \\
Swin-B/4W12-384~\cite{liu2021swin} & ImageNet & {\boldmath$70.41_{\pm0.41}$} & 
{\boldmath$45.06_{\pm0.02}$} & 
{\boldmath$69.83_{\pm0.37}$} & {\boldmath$70.41_{\pm0.42}$} & {\boldmath$69.69_{\pm0.46}$} \\
% \midrule
% Swin-CBM & ImageNet & 0.89 & 0.86 & 0.87 & 0.86 \\
\bottomrule
\end{tabular}
\end{table}

\begin{table}[t]
\vspace{-10pt}
\scriptsize
\centering
\captionsetup{font=scriptsize}
\caption{Performance comparison of model variants using the \texttt{Swin-B/4W12-384} backbone. The first two rows are baselines without a concept bottleneck (CB) layer, used for $8$-main class (MC) and $19$-subclass (SC) classification. The next four rows report CBMs trained with different concept sets: lesion descriptors ($47$), body parts ($49$), and both ($96$). The last merged rows present individual layer (SC \& MC) performance of a Hierarchical CBM (Type 1 \& 2) combining the CB layer with joint SC and MC classification as described in Fig ~\ref{fig:cbm_arc}. All metrics are results of end-to-end training and are averaged over $5$ random seeds.}
\label{tab:model_cbm}
\setlength{\tabcolsep}{4pt}
\begin{tabular}{r c c c c c c}
\toprule
\textbf{Classification head} & \textbf{Concepts} & \textbf{Accuracy} & \textbf{Precision} & \textbf{Sensitivity} & \textbf{F1 Score} & \textbf{Macro AUC}\\
\midrule
MC & - & {\boldmath$70.41_{\pm0.41}$} & {\boldmath$69.83_{\pm0.37}$} & {\boldmath$70.41_{\pm0.42}$} & {\boldmath$69.69_{\pm0.46}$} & {\boldmath$78.51_{\pm0.59}$}\\
SC & - & $58.27_{\pm0.22}$ & $56.64_{\pm0.45}$ & $58.27_{\pm0.22}$ & $56.81_{\pm0.43}$ & $83.11_{\pm2.55}$\\
\midrule
(CBM-D) Concepts + MC & Descriptors & $68.57_{\pm0.72}$ & $67.63_{\pm0.97}$ & $68.55_{\pm0.72}$ & $67.69_{\pm79.48}$ & $85.18_{\pm2.98}$\\
(CBM-B) Concepts + MC & Body parts & $68.38_{\pm0.31}$ & $67.69_{\pm0.51}$ & $68.31_{\pm0.27}$ & $67.90_{\pm0.26}$ & $84.96_{\pm1.27}$\\
Concepts + MC & Descriptors \& Body parts & $68.12_{\pm0.43}$ & $67.69_{\pm0.71}$ & $68.10_{\pm0.48}$ & $67.56_{\pm0.48}$ & $82.78_{\pm2.70}$\\
\midrule
(CBM-D) Concepts + SC & Descriptors & $56.42_{\pm0.01}$ & $55.72_{\pm0.01}$ & $56.51_{\pm0.01}$ & $55.63_{\pm0.01}$ & $80.16_{\pm0.01}$ \\
(CBM-B) Concepts + SC & Body parts & $55.88_{\pm0.01}$ & $55.47_{\pm0.01}$ & $55.90_{\pm0.01}$ & $54.87_{\pm0.01}$ & $78.94_{\pm0.01}$ \\
Concepts + SC & Descriptors \& Body parts &  $57.37_{\pm0.01}$ & $57.67_{\pm0.01}$ & $57.27_{\pm0.01}$ & $56.90_{\pm0.01}$ & $78.39_{\pm0.02}$ \\
\midrule
(Type1) SC & Descriptors \& Body parts & $53.98_{\pm0.40}$ & $56.12_{\pm0.74}$ & $53.98_{\pm0.39}$ & $54.49_{\pm0.66}$ & $76.13_{\pm1.52}$\\
(Type1) MC & Descriptors \& Body parts & $67.78_{\pm0.47}$    & $67.32_{\pm0.67}$ & $67.66_{\pm0.48}$ & $66.92_{\pm0.37}$  & $79.53_{\pm0.62}$\\
\hdashline
(Type 2) SC & Descriptors \& Body parts & $56.11_{\pm0.67}$ & $55.49_{\pm0.0.62}$ & $56.09_{\pm0.9}$ & $54.49_{\pm0.66}$ & $76.13_{\pm1.52}$\\
(Type 2) MC & Descriptors \& Body parts & $69.90_{\pm0.20}$    & $68.82_{\pm0.36}$ & $69.89_{\pm0.19}$ & $69.08_{\pm0.31}$  & $77.01_{\pm2.24}$\\
\bottomrule
\end{tabular}
\vspace{-10pt}
\end{table}
\subsection{Standard architectures}
DermaCon-IN comprises high-resolution clinical photographs with multiple co-occurring lesions, varied anatomical regions, and hierarchically structured multi-label annotations. We selected architectures based on their complementary modelling strengths to benchmark model performance under these conditions. Convolutional neural networks such as ResNet50~\cite{he2016deep}, DenseNet121~\cite{huang2017densely}, and EfficientNet (\texttt{EffNet-B4})~\cite{tan2020efficientnet} are effective in capturing localised texture patterns and edge-level features, owing to their convolutional inductive biases and limited receptive fields. To complement this, we incorporated Vision Transformer (ViT) architectures~\cite{dosovitskiy2020image}, which leverage self-attention to relate spatially distant regions within an image. The ViT variants evaluated includes ViT-Base (\texttt{ViT-B/16-224}~\cite{dosovitskiy2020image}, \texttt{ViT-B/16-384})~\cite{dosovitskiy2020image}, MaxViT-Base (\texttt{MaxViT-B/512})~\cite{tu2022maxvit}, and Swin-Base (\texttt{Swin-B/4W12-384})~\cite{liu2021swin}. Among these, the Swin Transformer consistently achieved the best results for Main class prediction across evaluation metrics, demonstrating improved handling of both multi-class classification and class imbalance. Table~\ref{tab:model_benchmark} summarises the performance of all models considered. Swin Transformer’s shifted window mechanism enables efficient modeling of non-contiguous regions, while its hierarchical representation captures both fine-grained lesion details and broader spatial patterns. These traits align closely with our dataset. We believe this alignment contributed to Swin’s better performance.

Based on these observations, we adapted the \texttt{\texttt{Swin-B/4W12-384}}~\cite{liu2021swin} variant of the Swin Transformer as the backbone for subsequent analysis and in concept bottleneck models (CBMs)~\cite{Koh2020Concept} as shown in Table~\ref{tab:model_cbm}. The model was initialised with weights pretrained on ImageNet-22k and fine-tuned end-to-end on our dataset. Input images were resized and padded to $512 \times 512$ for further classification. We performed a stratified, subject-wise 80:20~\label{data_split} split over Sub Class and reported all the results with the same split. We adapted weighted sampling strategies to handle class imbalance, and details of which are discussed in Supplementary Sec. B.

\subsection{Concept Bottleneck Modeling for Interpretablity}
\begin{wrapfigure}{r}{0.35\textwidth}
  \centering
  \vspace{-40pt}
\includegraphics[width=0.35\textwidth]{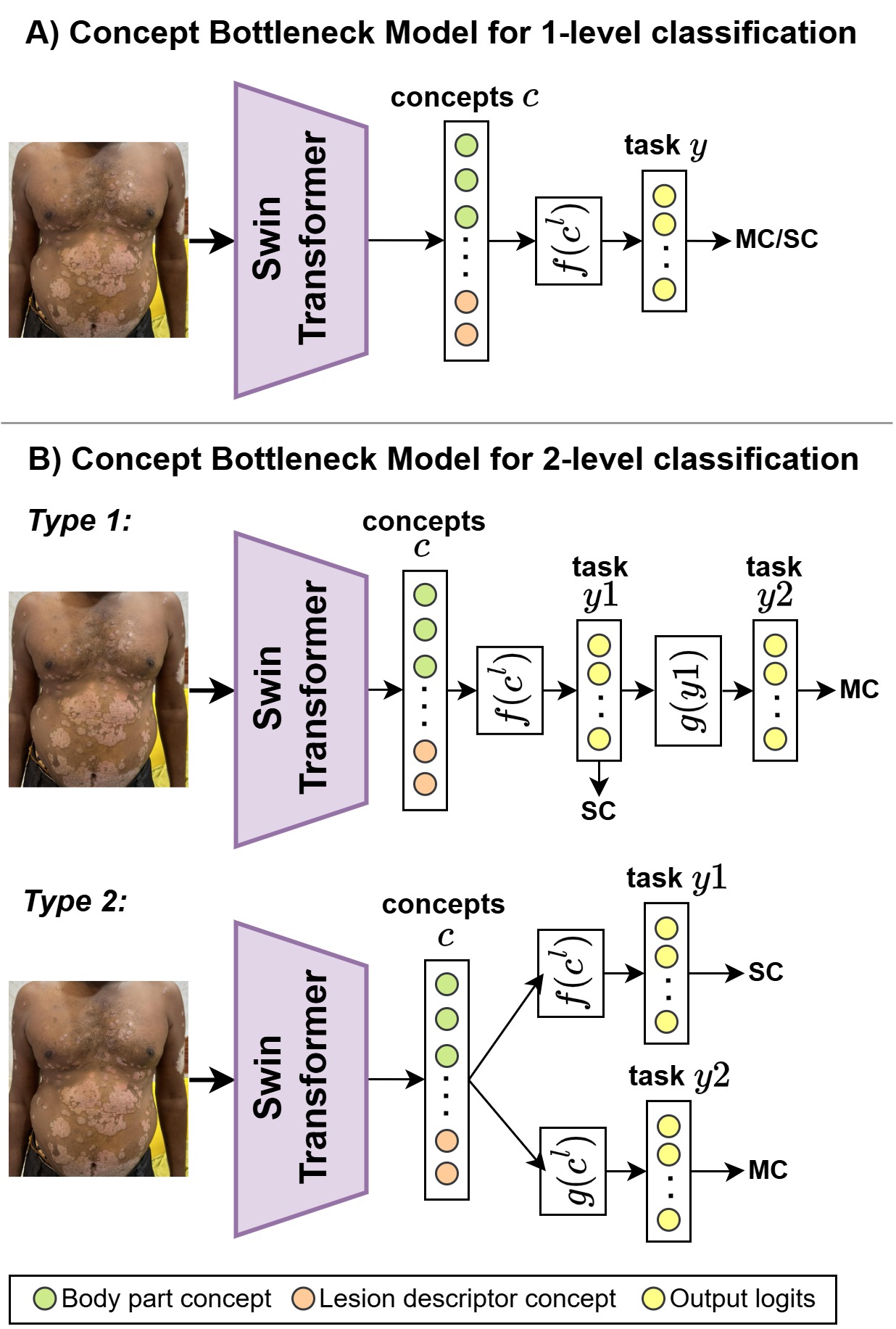}
  \vspace{-15pt}
  \captionsetup{font=scriptsize}
  \caption{Architectural setup of Concept-bottleneck models for Main class (MC) and Sub class (SC) classification}
    \vspace{-15pt}
  \label{fig:cbm_arc}
\end{wrapfigure}

\label{subsec:cbm}

\paragraph{Architecture.}
Given an input image \(x\), the Swin-Transformer encoder \(E_{\theta}\) yields a latent representation \(z = E_{\theta}(x)\). A linear projection maps \(z\) to \emph{concept logits} \(c^{\ell} \in \mathbb{R}^{B+D}\), which are then passed through a sigmoid activation to obtain the interpretable \emph{concept vector}
\(
c = [c^{\text{bp}},\,c^{\text{ld}}] = \sigma(c^{\ell}) \in [0,1]^{B+D},
\)
where \(c^{\text{bp}}\) denotes \(B\) \textbf{body-part} concepts and \(c^{\text{ld}}\) denotes \(D\) \textbf{lesion-descriptor} concepts. While both \(c\) and \(c^{\ell}\) are used for interpretability and concept supervision, the downstream classifier \(f\) operates on the concept logits \(c^{\ell}\) to produce task logits \(y = f(c^{\ell})\),
predicting either a \emph{Main-Class} (MC) or \emph{Sub-Class} (SC) label. This architecture is presented in Figure~\ref{fig:cbm_arc}(A).
\begin{wrapfigure}{r}{0.55\textwidth}
  \centering
  \vspace{-15pt}
\includegraphics[width=0.55\textwidth]{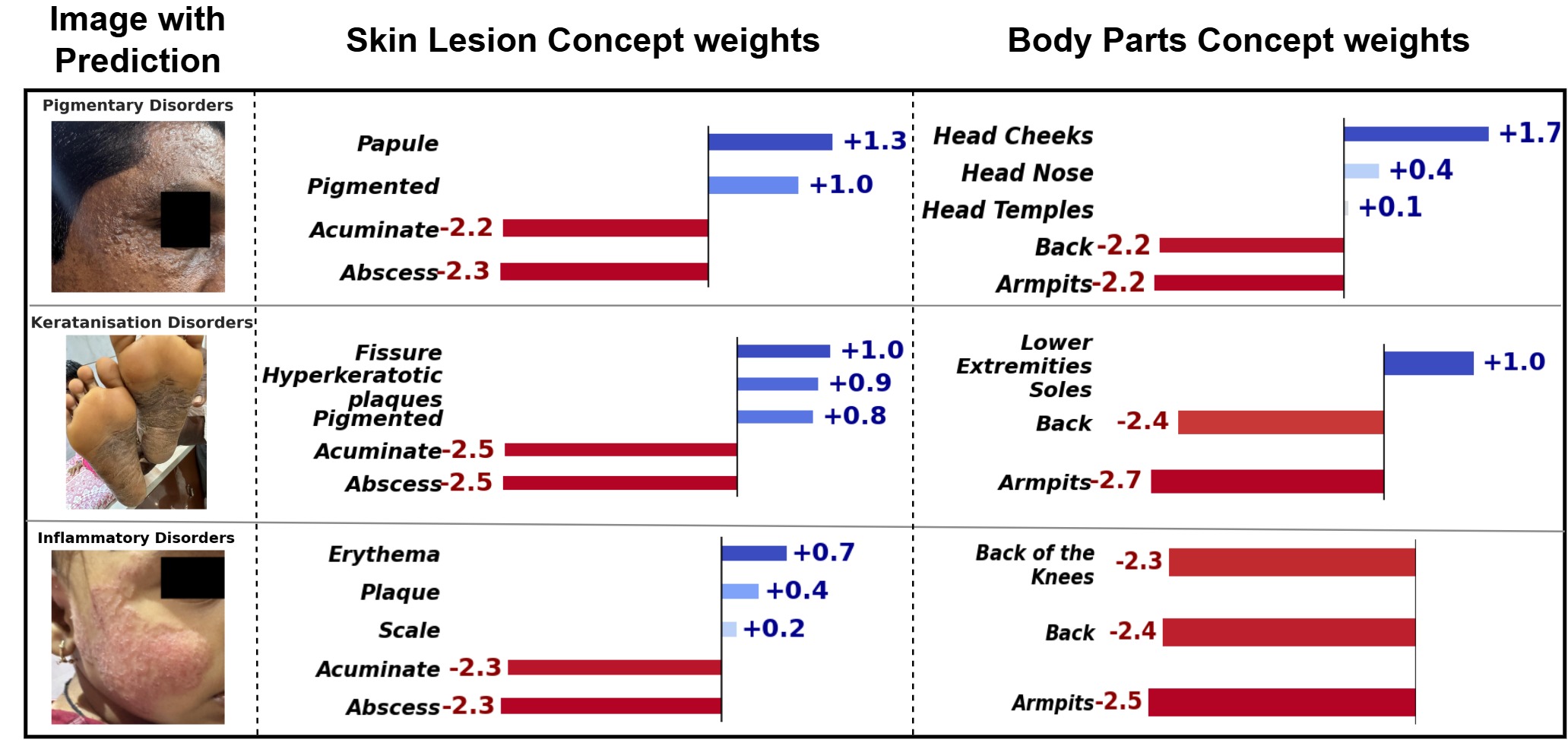}
  \vspace{-15pt}
  \captionsetup{font=scriptsize}
  \caption{Bar plot of top and bottom-k contributing concepts (lesion descriptors and body parts) for the model’s prediction. Contributions are shown as signed log-scaled weights derived from the CBM’s intermediate logits.}
    \vspace{-10pt}
  \label{fig:con_weight}
\end{wrapfigure}
\vspace{-15pt}
\paragraph{Concept ablation and joint–stream effects.}
To quantify the contribution of each concept group, we trained two ablated models:
\textbf{CBM-D}, which keeps only lesion-descriptor concepts
\(\smash{c^{\text{ld}}}\),
and
\textbf{CBM-B}, which keeps only body-part concepts
\(\smash{c^{\text{bp}}}\), both predicting MC labels.
Each single–stream variant retained high accuracy (Table~\ref {tab:model_benchmark}), demonstrating that the two concept families are independently learnable.

By contrast, the \emph{full} CBM represents the true clinical diagnostic fidelity, in which both
\(\smash{c^{\text{bp}}}\) and \(\smash{c^{\text{ld}}}\)
co-exist in the bottleneck, achieved performance comparable to the individual concept streams, but revealed a systematic imbalance in activation:
In many samples, only one concept group (typically descriptors) fired strongly, whereas the other (body parts) was under-activated (Fig.~\ref{fig:con_weight}, bottom row). This led to a modest but consistent drop in overall accuracy, pointing to a \emph{representational bottleneck} whereby competition for limited capacity biases the model toward a single semantic stream. These observations emphasize the need for improved multi-concept learning mechanisms that can balance several concept families simultaneously.

\vspace{-5pt}
\paragraph{Hierarchical CBMs.}
We explore two designs for joint SC–MC prediction (Figure~\ref {fig:cbm_arc}(B)):
\vspace{-5pt}
\begin{enumerate}
\item \textbf{Type 1 — cascade.}
Concepts first predict sub-classes via
\(y1 = f(c)\).
These logits are then mapped to the main classes through a second head
\(y2 = g(y1)\),
ensuring taxonomy consistency by construction.
\vspace{-4pt}
\item \textbf{Type 2 — parallel.}
Both SC and MC are predicted from the shared concept vector through
independent heads,
\(y1 = f(c)\) and
\(y2 = g(c)\),
leveraging multi-task learning for implicit regularization.
\end{enumerate}

\begin{wrapfigure}{r}{0.31\textwidth}
  \centering
  \vspace{-20pt}
\includegraphics[width=0.31\textwidth]{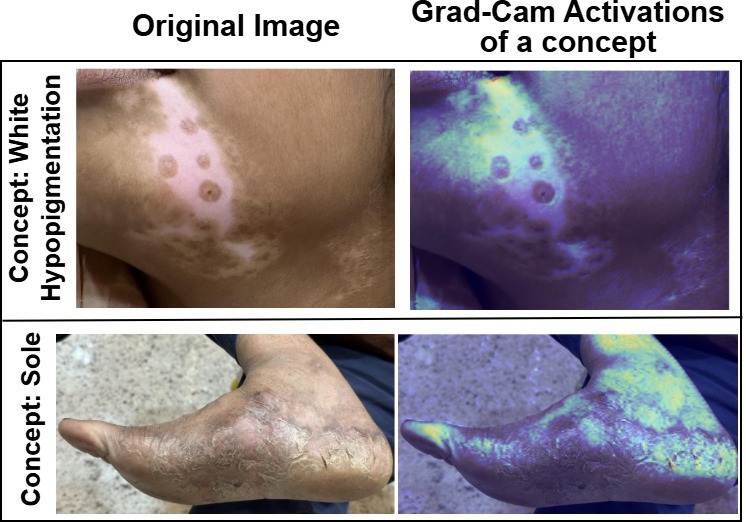}
  \vspace{-10pt}
  \captionsetup{font=scriptsize}
  \caption{Examples of Grad-cam visualizations over Swin Transformer by choosing a specific concept with the best CBM model (Type-2).}
    \vspace{-15pt}
  \label{fig:gradcam}
\end{wrapfigure}

Empirically, the \emph{parallel} configuration surpassed the
\emph{cascade} alternative across all evaluation metrics
(Table.~\ref{tab:model_benchmark}), likely due to effective regularization and information sharing through the multi-task learning setup.

\vspace{-5pt}
\paragraph{Qualitative Analysis.}
\begin{wrapfigure}{r}{0.5\textwidth}
  \centering
  \vspace{-10pt}
\includegraphics[width=0.5\textwidth]{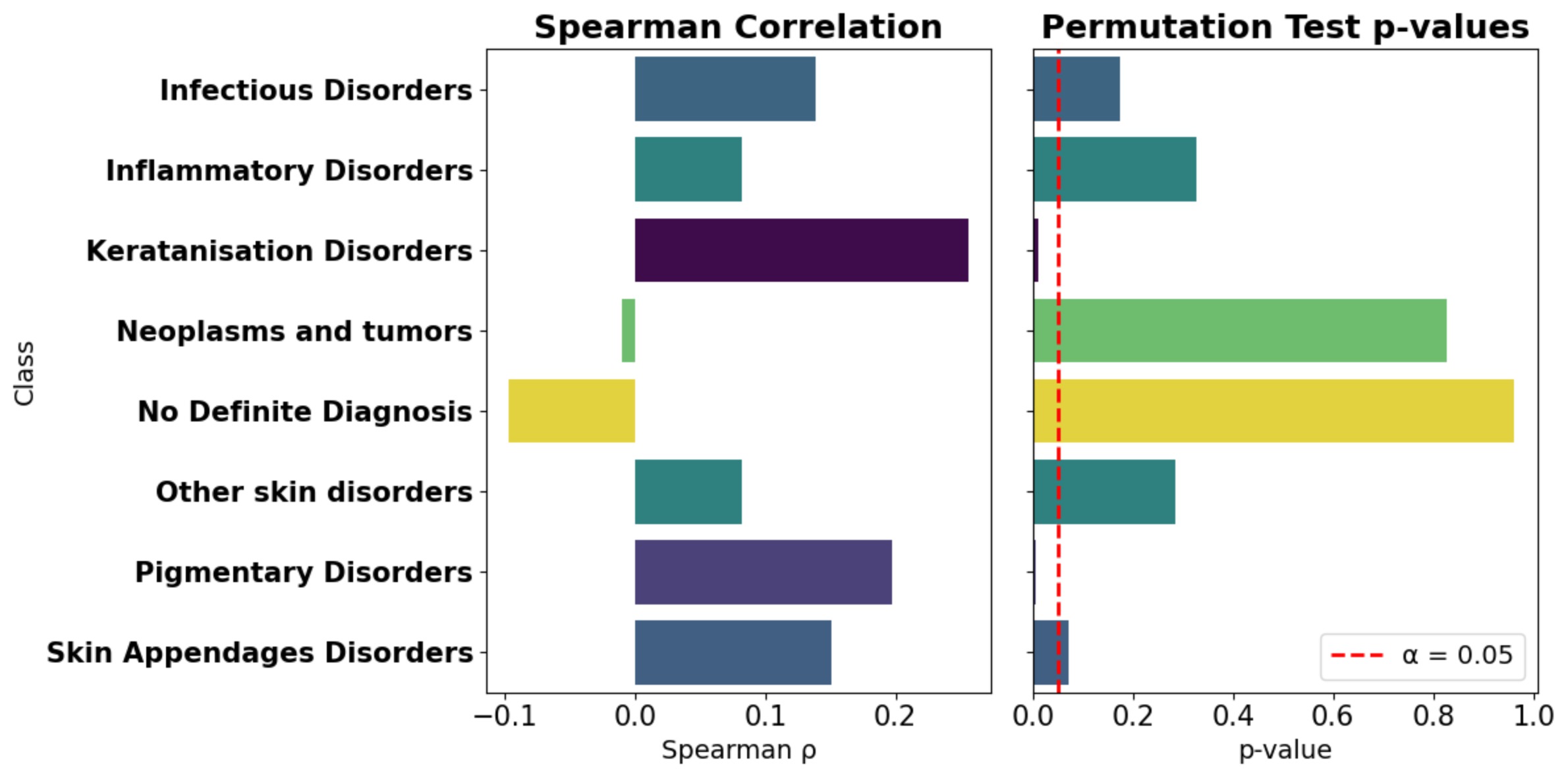}
  \vspace{-15pt}
  \captionsetup{font=scriptsize}
  \caption{Class-wise assessment of alignment between the model’s learned label weights over concepts and the dataset-derived correlation of concepts with class labels. The Spearman correlation measures the rank agreement between model-assigned concept importance and empirical concept–class correlations. Permutation test p-values (based on Pearson correlation) assess the statistical significance of this alignment.}
    \vspace{-10pt}
  \label{fig:spearman}
\end{wrapfigure}
To validate the spatial grounding of concepts, we employed Grad-CAM visualizations over Swin ViT on specific concept heads (Figure~\ref{fig:gradcam}). For each selected concept (from both descriptor and body part categories), we backpropagated gradients from the concept prediction to the image space, producing activation heatmaps. These visualizations confirmed that the model's concept activations were often localized to semantically and anatomically appropriate regions, supporting the faithfulness of the learned representations.

To further assess whether the model’s predictions are semantically grounded, we evaluate the alignment between its learned class-concept weights (MC branch of Type-2) and the statistical relevance (Pearson correlation, Section~\ref{sec:datset_overview}) of each concept to the class labels in the dataset (Figure~\ref{fig:spearman}). Alignment varies notably across diagnostic categories. \textit{Pigmentary Disorders} and \textit{Keratinisation Disorders} show strong Spearman correlations and statistically significant p-values ($p < 0.05$), suggesting that the model reliably prioritises clinically meaningful features for these classes. In contrast, \textit{Neoplasms and Tumors} show weak or negative alignment, indicating reliance on non-semantic or latent cues, possibly due to low representation in the dataset. Whereas \textit{No Definite Diagnosis} shows negative correlations as desired. Classes like \textit{Skin Appendageal Disorders, etc.} exhibit partial alignment.

Overall, these findings highlight that while the model is capable of learning semantically meaningful representations in some contexts, its reliance on concept supervision is uneven and class-dependent. This motivates future work in enforcing more robust concept--class alignment, particularly for clinically ambiguous or visually heterogeneous disease categories.

\section{Conclusion and Future Directions.} ~\label{conclusion and future direction} 
DermaCon-IN captures real-world dermatological presentations from the South Indian population, offering concept-level annotations such as body parts and disease descriptors. It addresses gaps in skin tone diversity and regional disease patterns, serving both as a region-specific benchmark and a valuable addition to global dermatological datasets. By enabling evaluation beyond labels, it supports clinically grounded, interpretable learning, a step closer towards clinical deployment.

\textbf{Targeted future work.}
Our findings point to several \emph{method-driven} avenues for closing the identified gaps. These steps can help models to not only classify accurately but also reason in clinically meaningful ways. They are as follows:
\vspace{-5pt}
\begin{itemize}
\item \textbf{Concept-weight normalisation:}
Adding explicit regularisers that penalise disproportionate reliance on a single concept group, encouraging balanced gradient flow.
\item \textbf{Hierarchy-consistent objectives:}
coupling the SC and MC heads with cross-level consistency losses to discourage contradictory evidence pathways.
\item \textbf{Curriculum and re-sampling strategies:}
oversampling under-represented concept combinations (e.g.\ body-part signals within neoplasm classes) to equalize learning pressure across concept space.
\end{itemize}

\vspace{-5pt}
\textbf{Broader integration.}
Because DermaCon-IN offers strong coverage of South-Indian skin tones and disease spectra, its concept annotations complement existing global datasets. Merging these resources will enable training of larger, more broadly
representative models, paving the way toward a unified \emph{foundation-level} representation for dermatological imaging. Looking ahead, we will expand DermaCon-IN as versioned releases sourced from additional centers across India, thereby deepening geographic, phenotypic, and skin-tone coverage while preserving curation standards. In parallel, we plan to add pixel-level lesion masks for a subset of images, enabling segmentation and tighter alignment between concepts and spatial evidence. We hope this resource will move dermatology AI closer to responsible clinical use.

\paragraph{Ethical Clearance Statement:}~\label{ethical clearance}
The dataset was collected in accordance with institutional ethical guidelines and has been approved by the Institute Ethics Committee of the Indian Institute of Technology Hyderabad under protocol number \textit{IITH/IEC/2025/01/05}.

\vspace{-5pt}
\paragraph{Implementation:}\label{compute} The experiments were run on 4 GPUs of Nvidia-A6000, each of $42$GB RAM. The dataset sizes around $\sim\!4$ GB. The code used in this work is available at \href{https://github.com/shan4035/DermaCon-IN-code.git}{GitHub}. 

\vspace{-5pt}
\paragraph{Availability and Licensing:}\label{data_availability_harvard_datverse}
The dataset can be downloaded at \href{https://dataverse.harvard.edu/dataset.xhtml?persistentId=doi:10.7910/DVN/W7OUZM}{Harvard Dataverse}. This work is licensed under CC BY-NC-SA 4.0. To view a copy of this license, visit \href{https://creativecommons.org/licenses/by-nc-sa/4.0/}{creativecommons.org}.

\vspace{-5pt}
\paragraph{Funding and Acknowledgement:}
This work was not supported by any external funding. All contributors were involved out of self-motivation and shared interest in advancing dermatological AI research. 

{\small
\bibliography{ref}  % do not include .bib extension
}
%%%%%%%%%%%%%%%%%%%%%%%%%%%%%%%%%%%%%%%%%%%%%%%%%%%%%%%%%%%%

\appendix

% \maketitle
\appendix
\cleardoublepage
\phantomsection
\addcontentsline{toc}{chapter}{Supplementary Material}
% \section*{Supplementary Material for DermaCon-IN}
\vspace*{2em}
% \adl@hline{<row-height>}{<dash-length>}
\begin{center}
    \LARGE \textbf{Supplementary Material for DermaCon-IN}
\end{center}
% \adl@hline{<row-height>}{<dash-length>}
\vspace{4em}

% % Custom TOC for Appendix
% \renewcommand{\cftsecleader}{\cftdotfill{\cftdotsep}}  % dotted lines
% % \vspace{0.5em}
% {
% \hypersetup{linkcolor=blue}
% \tableofcontents
% }
\vspace{-10pt}
% \tableofcontents

\section{Dataset Details}

\subsection{Rook’s Classification for a Hierarchical Framework}

% \paragraph{Limitations of Existing Datasets:}
Most dermatology datasets, such as \textit{SD-198}~\cite{sun2016benchmark}and \textit{SCIN}~\cite{ward2024creating}, employ flat diagnostic label structures without an overarching clinical taxonomy. Others, like \textit{Fitzpatrick17k}~\cite{groh2021evaluating} and \textit{SkinCon}~\cite{daneshjou2022skincon}, offer fine-grained or concept-level annotations but do not embed these within an etiologically structured or pathophysiology-aware hierarchy. As a result, they fall short of modeling the layered reasoning typical of clinical diagnosis. While sufficient for benchmarking classification models, these taxonomies lack clinical depth and fail to represent the structured reasoning used in dermatological diagnosis.

\paragraph{Clinical Relevance:}
Rook’s taxonomy~\cite{Griffiths2024Rook} introduces a hierarchical  framework grounded in disease etiology and pathological processes, organizing conditions into superclasses such as \textit{infectious}, \textit{inflammatory}, \textit{pigmentary}, and \textit{appendageal}, etc. This mirrors how clinicians formulate differential diagnoses from broad mechanisms to specific entities and enables AI systems to produce outputs that resonate with clinical logic.

\paragraph{Contextual Suitability:}
In Indian outpatient dermatology, where fungal, pigmentary, and inflammatory conditions predominate, this structure offers strong alignment with real-world case loads. Unlike neoplasm-focused Western datasets, Rook's~\cite {Griffiths2024Rook} schema reflects the diversity and prevalence of diseases encountered in Low \& Middle Income Countries (LMIC) settings, while remaining extensible to global contexts.

\paragraph{Curation and Learning Advantages:}
A hierarchical setup facilitates consistent labeling, balanced representation across superclasses, and structured sampling strategies. It also supports coarse-to-fine prediction models, multi-task learning, and scalable dataset expansion.

\paragraph{Interoperability with Clinical Systems:}
Crucially, Rook’s classification aligns with international standards such as the WHO’s ICD-10~\cite{who2019icd11} and ICD-11, which also structure diseases by etiology and anatomical relevance. Since ICD codes underpin electronic health records, billing, and decision support tools, adopting a Rook-aligned hierarchy ensures that the dataset remains interoperable with clinical infrastructures. This enables seamless integration into healthcare workflows, extending its utility well beyond academic modeling to real-world deployment.

\subsection{Descriptor Design and Clinical Alignment}

\paragraph{Descriptor Design and Selection:}
We annotate $5{,}450$ dermatological images using $47$ carefully selected descriptors encompassing primary lesions, secondary changes, pigmentary alterations, vascular anomalies, and surface/nail abnormalities. The taxonomy is rooted in established dermatological frameworks~\cite{bolognia2017dermatology}, ensuring clinical interpretability and consistency with medical curricula and diagnostic practice. While many descriptors overlap with those used in SkinCon~\cite{daneshjou2022skincon} (e.g., \textit{papule}, \textit{plaque}, \textit{vesicle}), our set incorporates necessary refinements based on regional prevalence and diagnostic relevance. Descriptors with high clinical utility but low frequency (e.g., \textit{poikiloderma}, \textit{gray}, \textit{salmon}) were retained to preserve diagnostic granularity. The descriptors such as \textit{translucent}, \textit{friable}, and \textit{dome-shaped} lacked distinct visual features in $2D$ clinical images and so were not included in the list of descriptors, which were otherwise part of SkinCon descriptors.

\paragraph{Contextual Tailoring for Indian Dermatology:}
Our descriptor set is designed with sensitivity to Indian outpatient dermatology. Conditions such as vitiligo and post-inflammatory hyperpigmentation necessitate precise pigmentary annotations (\textit{white}, \textit{gray}, \textit{brown}, \textit{pigmented}). Similarly, endemic infections (e.g., scabies, impetigo) justify the inclusion of \textit{crust}, \textit{burrow}, and \textit{abscess}. Chronic inflammatory presentations (\textit{lichenification}, \textit{hyperkeratotic plaques}) and nail findings (\textit{pitted nail}, \textit{discolored nail}) are also integrated based on their diagnostic frequency and clinical relevance in South Asia.

\paragraph{Standardization and Future Compatibility:}
Despite its regional grounding, our descriptor vocabulary remains aligned with global dermatological standards, enabling seamless integration with future datasets developed under similar clinically guided frameworks. By leveraging literature-backed terminology, our schema offers both backward compatibility with existing datasets like SkinCon~\cite{daneshjou2022skincon} and forward compatibility with evolving multimodal dermatology benchmarks.

\paragraph{Summary.}
Our descriptor design balances regional specificity with standardization. It enhances diagnostic interpretability, improves label quality for model training, and facilitates interoperability across future dermatological datasets. A detailed table of descriptors, definitions, and rationale is provided in Sup. Table~\ref{sec:descriptor_table}.

{\scriptsize
\begin{longtable}{r p{1.8cm} p{5cm} p{5cm}}
\caption{Complete Concept Descriptor Table with Definitions and Reasoning.} \\
\label{sec:descriptor_table} \\
\toprule
\textbf{S.No} & \textbf{Concept} & \textbf{Explaination} & \textbf{Reason for Choosing the Descriptor} \\
\midrule
\endfirsthead

\caption[]{Complete Concept Descriptor Table (continued)} \\
\toprule
\textbf{S.No} & \textbf{Concept} & \textbf{Explaination} & \textbf{Reason for Choosing the Descriptor} \\
\midrule
\endhead

\midrule
\multicolumn{4}{r}{{Continued on next page}} \\
\midrule
\endfoot

\bottomrule
\endlastfoot

1 & Abscess & A localized collection of pus within the dermis or subcutaneous tissue, typically surrounded by inflamed tissue; clinically presents as a painful, erythematous, fluctuant nodule. & Common in bacterial infections; present frequently in tropical and humid climates. \\
2 & Acuminate & A lesion with a tapering, pointed shape; commonly seen in viral warts. & Seen in viral warts and common sexually transmitted infections like condyloma in Indian OPDs. \\
3 & Atrophy &A reduction or thinning of tissue, which may involve the epidermal (outer skin layer), dermal (middle layer), or subcutaneous (deep fat and connective tissue layer). & Frequently observed in chronic corticosteroid use and dermatoses like lichen sclerosus. \\
% 3 & Black & Describes lesions with dense melanin or necrotic tissue; seen in melanomas, eschars. & Helps identify necrotic lesions or melanotic conditions in pigmented skin. \\
4 & Brown (Hyperpigmentation) & Darkened area of the skin due to excess melanin. & Extremely common in post-inflammatory states and melasma in the Indian population. \\
5 & Bulla & A circumscribed lesion > 1 cm in diameter that contains liquid (clear, serous or haemorrhagic). & Seen in autoimmune and infective blistering disorders; important in differential diagnosis. \\
6 & Burrow & It is a thin, wavy, slightly raised line on the skin, typically found between the fingers or toes, caused by the scabies mite burrowing under the surface.
& Highly relevant in scabies, which has a significant endemic prevalence in India. \\
7 & Comedo & Comedo is a blocked skin pore caused by the accumulation of oil and dead skin cells. An open comedo, commonly known as a blackhead, has an exposed surface where the trapped material darkens due to air exposure. A closed comedo, or whitehead, remains sealed under the skin, forming a small bump. & Key feature in acne, a major dermatological complaint in adolescents and young adults. \\
8 & Crust & Dried serum, blood or pus on the surface of the skin. & Present in infected eczemas and impetigo, common in Indian children. \\
9 & Cyst & Cyst is a papule or nodule that contains fluid or semi-fluid material, making it soft and fluctuant to touch. & Frequent presentation in both cosmetic and inflammatory conditions, like epidermoid cysts. \\
10 & Dilated Vein & Enlarged, visible superficial vein; may be seen in varicosities or venous insufficiency. & Observed in varicose conditions and vascular anomalies, especially in rural patients. \\
11 & Discolored Nail & Alteration in nail color due to trauma, infection, pigmentation, or systemic disease. & an Indicator of systemic illness, fungal infections, or trauma-related disorders. \\
12 & Edema & Swelling of the skin due to fluid accumulation in the dermis or subcutaneous tissue. & Secondary signs in infections, inflammatory conditions, and systemic diseases. \\
13 & Erosion & Superficial loss of the epidermis, often due to ruptured blisters; heals without scarring. & Occurs in healing blisters and vesiculobullous diseases seen in Indian settings. \\
14 & Erythema & Redness of the skin due to increased blood supply. & Fundamental indicator of inflammation; often subtle in pigmented skin. \\
15 & Excoriation & A loss of the epidermis and a portion of the dermis due to scratching or an exogenous injury. & Seen in pruritic disorders like scabies, atopic dermatitis, and lichen simplex chronicus. \\
16 & Exophytic/Fungating & It describes a type of lesion that grows rapidly, breaks through the skin surface, and often appears ulcerated, foul-smelling, and infected, resembling a fungus-like mass. & Helps classify malignant and advanced skin lesions; relevant in tertiary care. \\
17 & Exudate & Oozing fluid composed of serum, pus, or blood; typically due to inflammation or infection. & Observed in infected wounds, ulcers, and pyodermas. \\
18 & Fissure & Fissure is a linear crack or break in the skin that begins in the outermost layer (stratum corneum) and may extend into the deeper dermis, often causing pain or bleeding. & Very common in xerotic skin conditions and hand/foot eczema exacerbated by occupational exposure. \\
19 & Flat-topped & Lesion with a flattened horizontal surface; characteristic of lichen planus. & Key for diagnosing lichen planus, prevalent in Indian adults. \\
20 & Gray & A color descriptor typically indicating post-inflammatory hyperpigmentation or pigment incontinence. & Represents pigment incontinence or deeper melanin deposition; seen in pigmented disorders. \\
21 & Hair Patch & Hair patch refers to a localized area on the skin where hair is either abnormally present (increased density or unusual location) or absent (loss of hair in a defined area) & Helps identify abnormal loss or growth of hair in a specific area. \\
22 & Hyperkeratotic plaques & Thickened plaques with an excessive build-up of keratin. & Observed in psoriasis, lichen simplex, and chronic eczema, common in Indian clinics. \\
23 & Induration & It refers to an area of the skin or tissue that feels firm or hardened to the touch, without any underlying calcification or bone formation. & Helps differentiate infections (e.g., cellulitis) or granulomatous conditions (e.g., leprosy). \\
24 & Lichenification & It is thickened, rough skin with exaggerated skin lines, typically resulting from repeated rubbing or scratching& Common in chronic atopic and lichen simplex; consequence of habitual scratching. \\
25 & Macule & A flat, circumscribed, nonpalpable lesion that differs in colour from the surrounding skin. & Key to diagnosing pigmentary and vascular disorders like vitiligo and leprosy. \\
26 & Nodule & An elevated, solid, palpable lesion > 1 cm usually located primarily in the dermis and/or subcutis. & Important for deep fungal infections, cutaneous TB, or cystic swellings in endemic areas. \\
27 & Papule & An elevated, solid, palpable lesion that is $\leq$ 1 cm in diameter. & Seen in common dermatoses like folliculitis, acne, and viral warts. \\
28 & Patch & A large area of colour change, with a smooth surface. & Central to identifying vitiligo, pityriasis alba, and leprosy. \\
29 & Pedunculated & Lesion attached by a narrow stalk. & Helps in the classification of benign tumors like acrochordons or neurofibromas. \\
30 & Pigmented & Lesions exhibiting increased pigment; may be brown, gray, or black. & Essential in differentiating dermatoses on brown skin; pigmentary presentations dominate. \\
31 & Pitted Nail & Small depressions on the nail surface. & Common signs of nail psoriasis and alopecia areata. \\
32 & Plaque & A circumscribed, palpable lesion $\geq$ 1 cm in diameter; most plaques are elevated. & Describes major lesion morphology in tinea, psoriasis, and lichen simplex. \\
33 & Poikiloderma & Simultaneous presence of atrophy, telangiectasia and hypo and hyperpigmentation. & Seen in late-stage connective tissue diseases; needs documentation in atypical Indian cases. \\
34 & Purpura/Petechiae & Haemorrhage into the skin due to pathological processes, primarily of blood vessels. & Observed in vasculitis and hematological disorders. \\
35 & Pustule & A circumscribed lesion that contains pus. & Seen in acne, folliculitis, and impetigo; frequently present in outpatient cases. \\
36 & Salmon & Pink-orange hue used to describe psoriatic lesions, especially on lighter skin tones. & Color reference for certain psoriatic plaques in lighter Indian skin tones. \\
37 & Scale & A visible accumulation of keratin, forming a flat plate or flake. & Typical in dermatophytosis, psoriasis, and seborrheic dermatitis; common in humid climates. \\
38 & Scar & Fibrotic replacement of normal skin architecture after injury. & Important to track disease healing and secondary changes post-injury or intervention. \\
39 & Striae & Linear atrophic lesions due to dermal tearing. & Common due to corticosteroid use, obesity, puberty, and pregnancy-related changes. \\
40 & Telangiectasia & Permanently dilated capillaries. & Seen in rosacea, lupus, and long-term corticosteroid use. \\
41 & Ulcer & Full-thickness loss of the epidermis plus at least a portion of the dermis. & Crucial for identifying diabetic foot, leprosy, and chronic venous ulcers. \\
42 & Vesicle & A circumscribed lesion $\leq$ 1 cm in diameter that contains liquid (clear, serous or haemorrhagic). & Key feature in varicella, dermatitis herpetiformis, and contact dermatitis. \\
43 & Warty & Verrucous surface resembling a wart; rough and irregular. & Descriptive of HPV-induced lesions and seborrheic keratoses, which are common in the elderly in India. \\
44 & Wheal & A transient elevation of the skin due to dermal edema. & Seen in urticaria due to infections, drugs, and food reactions. \\
45 & White (Hypopigmentation) & Lighter than normal skin color due to loss or reduction of melanin. & Central to vitiligo, pityriasis alba, and tinea versicolor, which are frequent in India. \\
46 & Xerosis & Abnormal dryness of the skin. & Highly prevalent due to hygiene practices, hard water, and low humidity in winter. \\
47 & Yellow & Describes lesions with lipid, keratin, or bile pigment. & Seen in xanthomas, sebaceous discharge, and bacterial pustules. \\
\end{longtable}}

\subsection{Choice of Anatomical Site}
We developed our body region taxonomy by aligning it with how dermatologists reason through diagnoses in clinical settings. Standard references such as \textit{Rook’s Dermatology}~\cite{Griffiths2024Rook}, \textit{Fitzpatrick’s Dermatology}, and \textit{Bolognia’s Dermatology}~\cite{bolognia2017dermatology} consistently describe skin conditions based on their anatomical distribution. These texts emphasize that lesion location plays a central role in diagnosis, distinguishing, for example, mucosal from cutaneous presentations. By mirroring these clinically grounded patterns, we ensured that our descriptors reflect the spatial logic used in real-world diagnostic workflows.

% \paragraph{Integration with medical ontologies:}
% We aligned our descriptors with formal classification systems such as ICD-11~\cite{who2019icd11}, which assign region-specific diagnostic codes (e.g., L21.0 for seborrheic dermatitis of the scalp). This semantic compatibility enables linkage to electronic health records (EHRs),  and cross-mapping with ontology-driven datasets.

\paragraph{Coverage gaps in existing datasets:}
We reviewed widely used dermatology datasets—ISIC, Fitzpatrick17k~\cite{groh2021evaluating}, SD-198~\cite{sun2016benchmark}, etc., and found that they often lack precise anatomical context. Most provide cropped images that obscure lesion location or annotate broad categories like ``face'' or ``limb,'' limiting their clinical utility. To address this, we explicitly included underrepresented yet diagnostically critical regions such as the armpits and groin area. These regions are key to diagnosing conditions like candidiasis and tinea infections.

\paragraph{Hierarchical design for diagnostic reasoning:}
We designed our annotation hierarchy to support models that reason at multiple anatomical resolutions. Dermatologists often start with coarse region-based hypotheses and further refine them until morphology and context become clearer. Our taxonomy enables similar flexibility, allowing models to learn general patterns at macro levels while capturing fine-grained distinctions when needed. This structure mirrors how clinicians disambiguate conditions with overlapping visual features based on location.
The complete list of anatomical regions used in our annotation schema is provided in the Datasheet included with the supplementary material for reference.

\subsection{Comparative Coverage of Dermatological Disease Categories Across Public Datasets}
\label{subsec:dataset_comparison}
 Most publicly available dermatology datasets were developed for specific diagnostic tasks, often skin cancer triage, and do not reflect the full diagnostic spectrum seen in routine outpatient clinics, particularly in low and middle-income countries (LMICs) where poor maintenance of hygiene amongst the population is a key factor for the spread of infectious diseases. As a result, critical categories such as infectious disorders, pigmentary changes, and appendageal conditions are underrepresented or absent entirely. To assess where our dataset stands in relation to existing benchmarks, we compiled an exhaustive comparison in the Sup. Table~\ref{tab:exhaustive_comparison} of major dermatology datasets across clinically meaningful diagnostic categories.

\begin{table}[h]
\centering
\scriptsize
\caption{Comparative distribution (in \%) of dermatological disease categories across public datasets. Only our dataset provides a good representation across all categories as encountered in routine outpatient care, including mixed diagnoses.}
% \resizebox{\textwidth}{!}{%
\setlength{\tabcolsep}{1.5pt}
\begin{tabular}{lccccccc p{2.8cm}}
\toprule
\textbf{Dataset} & \textbf{Infectious} & \makecell{\textbf{Inflammatory} \\ \textbf{(incl. keratinisation)}} & \textbf{Pigmentary} & \makecell{\textbf{Appendageal} \\ \textbf{(Acne/Hair)}} & \textbf{Neoplastic} & \makecell{\textbf{No Definite} \\ \textbf{Diagnosis}} & \textbf{Others} & \textbf{Key Observations} \\
\midrule
\textbf{DermaCon-IN} & 40.86 & 30.53 & 15.25 & 8.81 & 0.95 & 0.68 & 2.92 & Full-spectrum, OPD-aligned dataset designed for LMIC clinical diversity \\
SCIN & 21.85 & 56.5 & 0.75 & Sparse & 5.20 & Not included & 15.0 & Inflammatory-heavy; lacks uncertain and appendageal representation \\
Fitzpatrick17k & -- & 65.67 & 7.2 & Sparse & 26.7 & Not included & -- & Inflammatory and neoplastic skew; lacks diagnostic uncertainty \\
PASSION & 63.52 & 25.05 & Sparse & -- & -- & Not included & 11.43 & Pediatric LMIC dataset; lacks neoplastic and appendageal coverage \\
DDI & 1.66 & 2.28 & 0.46 & Sparse & 92.7 & Not included & 3.0 & Biopsy-focused dataset; neoplasm-dominant \\
ISIC 2020 & -- & -- & -- & -- & 100.00 & Not included & -- & Exclusive neoplasm dataset \\
PH2 & -- & -- & -- & -- & 100.00 & Not included & -- & Exclusive neoplasm dataset \\
HAM10k & -- & -- & -- & -- & 100.00 & Not included & -- & Exclusive neoplasm dataset \\
PAD-UFES & -- & -- & -- & -- & 100.00 & Not included & -- & Cancer-focused dataset only \\
\bottomrule
\end{tabular}%
% }
\label{tab:exhaustive_comparison}
\end{table}

\subsection{Other stats of the DermaCon-IN dataset}
\begin{figure}
  \centering
  \vspace{-10pt}
  \includegraphics[width=\linewidth]{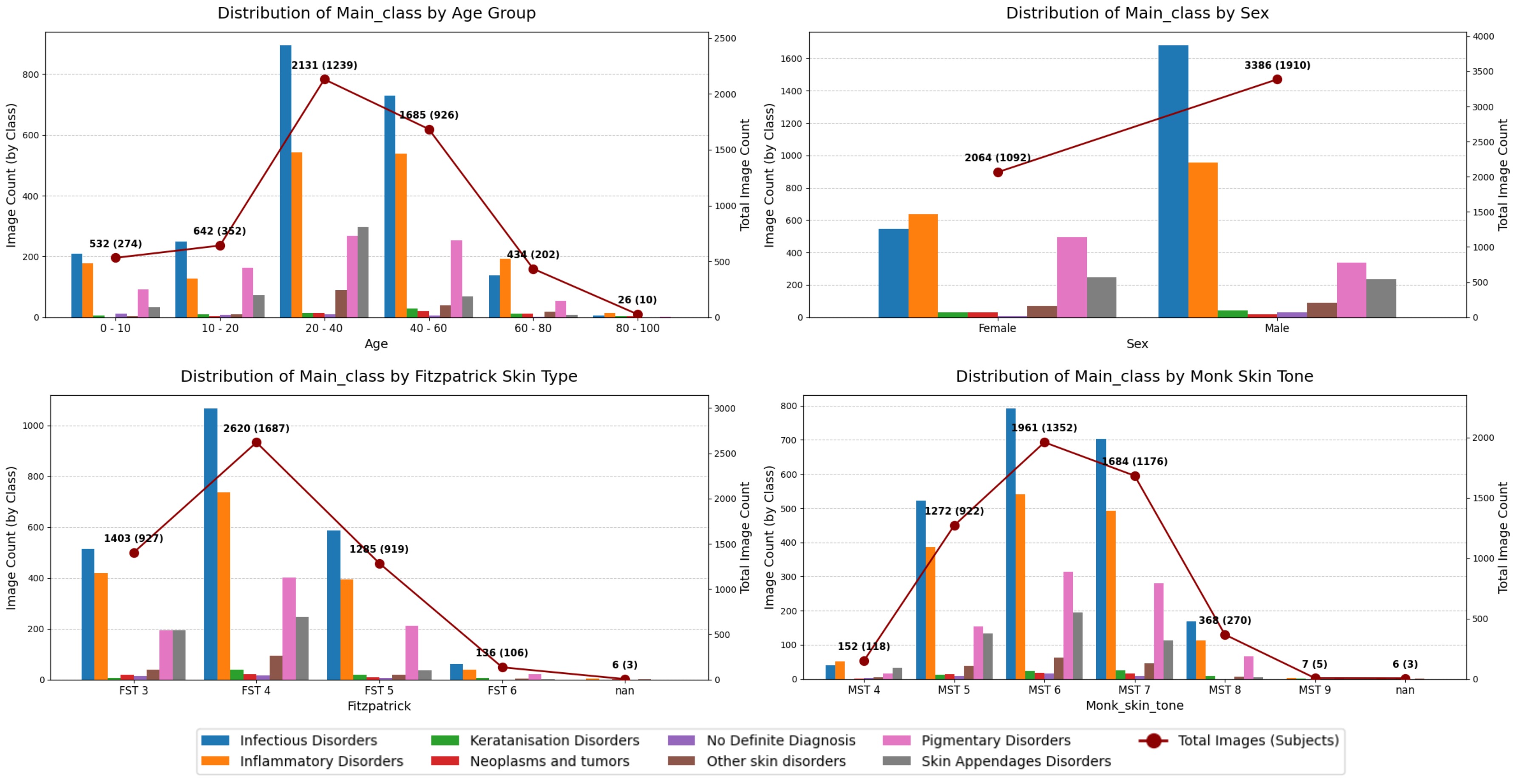}
  \vspace{-10pt}
  \caption{Plots for distribution of dermatological Main Classes (MC) across patient demographics. Each bar represents the number of images per category (Age group, Fitzpatrick type, Monk Skin Tone, or Sex), while the red line denotes the total number of images. Numeric annotations above each point indicate image count followed by subject count in parentheses.}
  % \vspace{-50pt}
  \label{fig:demographics}
\end{figure}

\paragraph{Demographic and Phenotypic Distribution:}
The dataset reveals distinct demographic and phenotypic trends across age, sex, and skin tone that shape its clinical composition (Sup. Figure~\ref{fig:demographics}). Most images are concentrated in the 20–40 age group, with secondary peaks in 10–20 and 40–60, reflecting outpatient demand among working-age individuals. Children (0–10) and older adults (60+) are relatively underrepresented. A notable sex imbalance exists, with males contributing more samples (3386 vs. 2064), possibly due to sociocultural factors. Infectious and pigmentary disorders dominate across both sexes, with females also showing higher counts for appendageal and inflammatory conditions.

In terms of skin tone, Fitzpatrick Types 4 and 5 are most prevalent, followed by Type 3. Type 6 is modestly represented (136 images), but is still less frequent relative to mid-tone categories. The Monk Skin Tone distribution similarly centers on MST 6 and 7, with lighter (MST 4) and darker tones (MST 8–9) less represented. Across all groups, infectious, inflammatory, and pigmentary disorders remain most common.

\begin{figure}
  \centering
  \includegraphics[width=\linewidth]{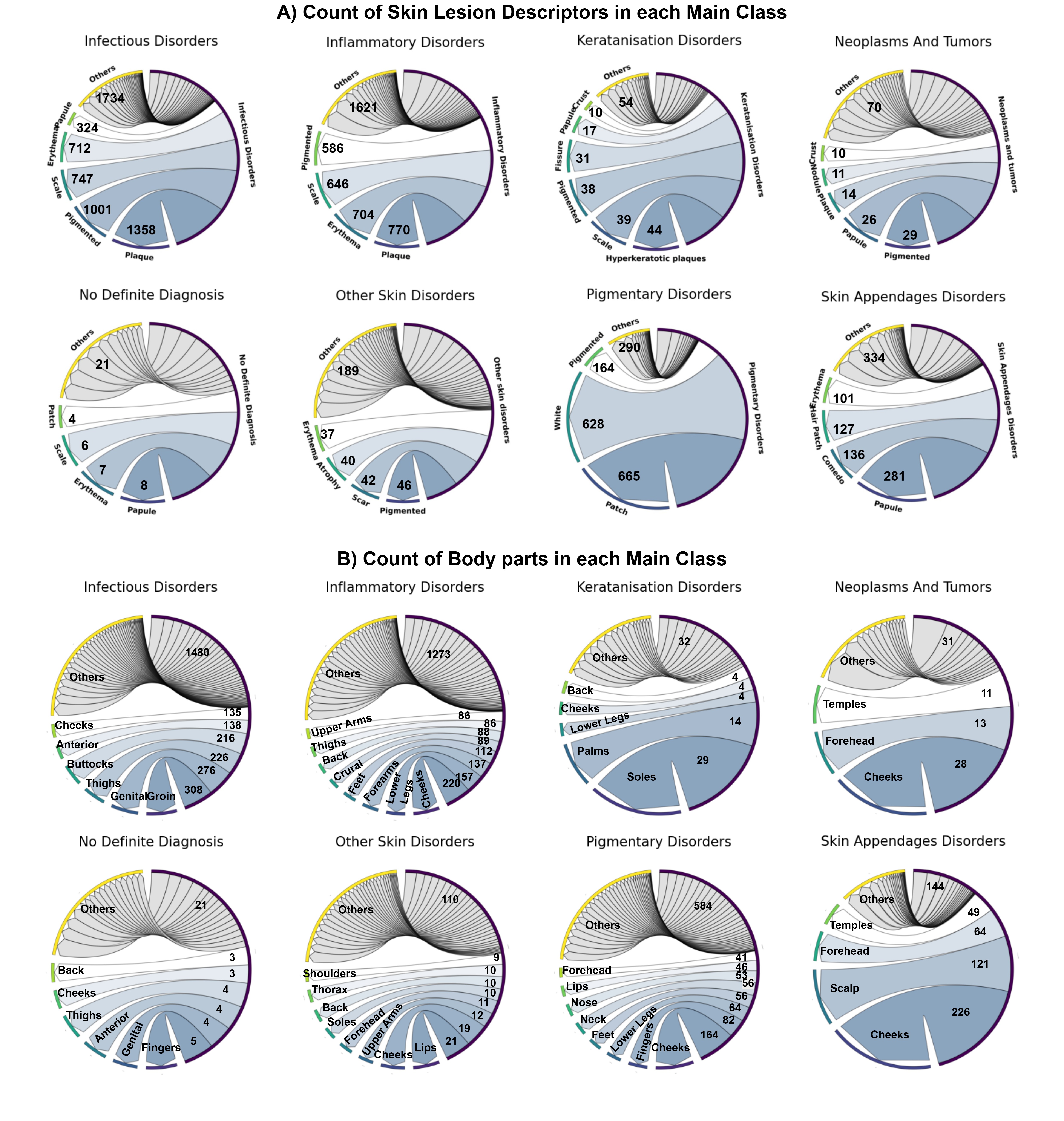}
  \caption{Circular plots of counts of Concepts (both skin lesion descriptors and body parts) for each main class skin labels.}
  \label{fig:con_count}
\end{figure}

\paragraph{Concept Distribution:} A detailed analysis of the concept counts per Main class is presented in the Sup. Fig.~\ref{fig:con_count}. The figure demonstrates the occurrence of shared concepts across the main class. Further, it also suggests an imbalance of concepts within each class.

\section{Model Architecture Details}

\subsection{Concept Bottleneck Modeling}
We investigate two modeling variants under the Concept Bottleneck framework: a 1-level CBM that performs either main class (MC) or sub-class (SC) prediction independently, and a 2-level CBM that jointly predicts both levels in a hierarchical structure. While both designs share a common concept encoder comprising a Swin Transformer followed by dropout and a linear projection to produce concept logits \( c^l \), they differ in their classification strategies as discussed below. Architectural and training hyperparameters for all CBM variants are summarized in Sup. Table~\ref{tab:cbm_hyperparams}.

\subsubsection{1-level CBM for MC/ SC Classification}
\label{sec:cbm_mc_only}
To explore the utility of medically grounded concepts in isolation, we train a Concept Bottleneck Model (CBM) that predicts dermatological categories, either main classes (MC) or sub-classes (SC), directly from concept representations. This model maps image features to concept logits and performs classification using those logits alone.

\paragraph{Model Architecture.}
An input image \( x \in \mathbb{R}^{3 \times H \times W} \) is first passed through a Swin Transformer backbone to extract a feature embedding. This embedding is then followed by a fully connected projection having Dropout, to produce a concept logit vector:
\[
c^l = \texttt{Linear}(\texttt{Dropout}(\texttt{SwinTransformer}(x))) \in \mathbb{R}^K
\]
The raw concept logits \( c^l \) are used as input to a linear classification head:
\[
y = f(c^l)
\]
where \( f(\cdot) \) is a single-layer classifier mapping to either subclass or main class logits depending on the training objective.

\paragraph{Training Objective.}
The model is trained with two primary losses:
\begin{align*}
\mathcal{L}_{\text{concept}} &= \text{BCEWithLogits}(c^l, c_{\text{true}}) \\
\mathcal{L}_{\text{class}} &= \text{CrossEntropy}(f(c^l), y_{\text{true}})
\end{align*}
The total loss is:
\[
\mathcal{L}_{\text{total}} = \mathcal{L}_{\text{concept}} + \mathcal{L}_{\text{class}} + \lambda \|c^l\|_1
\]
An L1 regularization term encourages sparsity in the concept representation, invoking the relevant concepts, improving interpretability, and acting as a regularizer during optimization.

\paragraph{Sampling and Imbalance Handling.}
To account for label imbalance in the main class distribution, we compute per-sample weights based on the inverse frequency of the main class and use a \texttt{WeightedRandomSampler} during training.

\vspace{5pt}
\begin{table}[h]
\centering
\scriptsize
\caption{Training and architecture hyperparameters for all CBM models.}
\label{tab:cbm_hyperparams}
\begin{tabular}{ll}
\toprule
\textbf{Component}              & \textbf{Value / Setting} \\ \midrule
Backbone architecture          & Swin-B (patch4, window12, 384) \\
Image resolution               & \( 512 \times 512 \) \\
Concept vector dimension \( K \) & 96 \\
Dropout before concept head    & 0.3 \\
Concept supervision loss       & BCEWithLogits (sigmoid applied internally) \\
Task heads \( f(\cdot), g(\cdot) \) & Linear layers (no hidden layer) \\
SC/MC loss                     & CrossEntropy with label smoothing \( \epsilon = 0.1 \) \\
L1 regularization weight \( \lambda \) & \( 1 \times 10^{-4} \) \\
Optimizer                      & AdamW \\
Learning rate                  & \( 3 \times 10^{-5} \) \\
Weight decay                   & \( 1 \times 10^{-4} \) \\
Scheduler                      & CosineAnnealingLR (T\textsubscript{max}=30, min LR=$10^{-6}$) \\
Training epochs                & 40 \\
Batch size                     & 130 (2-level), 100 (1-level) \\
Multi-GPU support              & DataParallel (4 GPUs) \\
\bottomrule
\end{tabular}
\end{table}

\subsubsection{2-level CBM for MC \& SC Classification}
\label{sec:cbm_formulation}

We propose a two-level hierarchical prediction framework based on Concept Bottleneck Models (CBMs), where disease predictions are structured via an interpretable intermediate layer of clinical concepts (refer Fig. 7 of the main paper). The task is to jointly predict a fine-grained dermatological \textit{sub-class} label \( y^1 \in \{1, \dots, N_{\text{sub}}\} \) and a coarser \textit{main class} label \( y^2 \in \{1, \dots, N_{\text{main}}\} \), from an input image \( x \in \mathbb{R}^{3 \times H \times W} \).

\vspace{5pt}
\paragraph{Concept Space.}
A vision backbone (Swin Transformer) maps the input image to a $K$-dimensional concept logit vector:
\[
c^l = \texttt{Linear}(\texttt{Dropout}(\texttt{SwinTransformer}(x))) \in \mathbb{R}^K
\]
We define \( c = \sigma(c^l) \in [0,1]^K \) as the sigmoid-activated concept probabilities, which are supervised using binary multi-label concept annotations. The raw logits \( c^l \), not the sigmoid outputs, are further used for downstream disease classification, preserving gradient flow and avoiding saturation effects.

\paragraph{Type 1: Cascade Architecture.}
This model enforces taxonomy consistency by chaining predictions through intermediate subclass logits:
\[
y^1 = f(c^l), \quad y^2 = g(y^1)
\]
Here, \( f(\cdot) \) is a linear layer mapping concept logits to subclass logits \( \in \mathbb{R}^{N_{\text{sub}}} \), and \( g(\cdot) \) is another linear layer mapping subclass logits to main class logits \( \in \mathbb{R}^{N_{\text{main}}} \). This design structurally enforces taxonomic consistency between sub- and main classes, aligning with medical hierarchies. However, performance is inherently constrained by the reliability of the first-stage prediction \( y^1 \), making it prone to error propagation.

\paragraph{Type 2: Parallel Architecture.}
Instead of sequential dependency, both SC and MC are predicted directly from concept logits:
\[
y^1 = f(c^l), \quad y^2 = g(c^l)
\]
Here, \( f(\cdot) \) and \( g(\cdot) \) are task-specific linear classifiers. This decouples the learning paths while maintaining shared semantic grounding via concepts. The architecture benefits from multi-task supervision and avoids dependency on intermediate task outputs. It allows the model to flexibly learn patterns that are specific to either task while still being grounded in a common, interpretable representation.

\vspace{5pt}
\paragraph{Loss Function.}
The overall objective for each image includes:
\begin{itemize}
    \item Binary cross-entropy loss on sigmoid-transformed concepts:
    \[
    \mathcal{L}_{\text{concept}} = \text{BCEWithLogits}(c^l, c_{\text{true}})
    \]
    \item Cross-entropy loss on subclass logits:
    \[
    \mathcal{L}_{\text{SC}} = \text{CrossEntropy}(f(c^l), y^1_{\text{true}})
    \]
    \item Cross-entropy loss on main class logits:
    \[
    \mathcal{L}_{\text{MC}} =
    \begin{cases}
    \text{CrossEntropy}(g(y^1), y^2_{\text{true}}), & \text{Type 1} \\
    \text{CrossEntropy}(g(c^l), y^2_{\text{true}}), & \text{Type 2}
    \end{cases}
    \]
    \item L1 regularization on concept logits to promote sparsity:
    \[
    \mathcal{L}_{\text{L1}} = \lambda \| c^l \|_1
    \]
\end{itemize}

The total loss is:
\[
\mathcal{L}_{\text{total}} = \mathcal{L}_{\text{concept}} + \mathcal{L}_{\text{SC}} + \mathcal{L}_{\text{MC}} + \mathcal{L}_{\text{L1}}
\]

\vspace{5pt}
\paragraph{Class Imbalance Handling.}
To account for imbalanced subclass and main class distributions, we assign sample-wise weights based on inverse frequency. For each sample \( i \), the final weight is computed as:
\[
w_i = \frac{1}{2} \left( \frac{1}{f(y^1_i)} + \frac{1}{f(y^2_i)} \right)
\]
where \( f(\cdot) \) is the empirical class frequency. These weights are used with a \texttt{WeightedRandomSampler} to ensure that rare classes contribute adequately during training.

\subsubsection{Handling Variable Image Resolutions.}
Dermatological images in real-world clinical datasets often exhibit heterogeneous resolutions and aspect ratios. To address this, we first resize images such that their longer side does not exceed 512 pixels while maintaining the aspect ratio. The resized image is then zero-padded to \(512 \times 512\) to ensure uniform input dimensions across batches. This approach preserves image content without distortion and enables efficient batch processing while maintaining compatibility with the fixed input size expected by the Swin Transformer backbone. Padding is handled dynamically during training and validation through a custom collate function.

\subsection{Detailed Performance Analysis}
\paragraph{Main Class Classification Performance.}
Sup. Table~\ref{tab:swin_mc_results} presents a detailed evaluation of our best-performing models for main class (MC) prediction. We report Top-1, Top-3, and Top-5 classification accuracies along with macro AUC values (\%) across all main classes and per-class breakdowns.

% The Swin-only model achieves a Top-1 accuracy of 70.88\% and a macro AUC of 78.36\%, with particularly strong performance on \textit{Skin Appendages} and \textit{Infectious} disorders. However, it underperforms significantly for classes with few examples such as \textit{Keratinisation Disorders}, \textit{Neoplasms}, and \textit{No Definite Diagnosis}.

% The Concept Bottleneck Model (CBM), trained with concept supervision and main class labels, consistently improves upon Swin across all metrics. It records a +3--5\% improvement in Top-1 accuracy and macro AUC, highlighting the benefit of intermediate semantic representations. Notably, Top-5 accuracy improves to 96.57\%, and AUC increases to 79.17\%.
For example, in inflammatory and pigmentary disorders, Top-1 accuracies range between 65–73\%, but Top-5 scores consistently rise above 95\%, indicating that the correct class is almost always among the top few ranked predictions. This pattern holds across infectious and appendageal disorders as well, with Top-5 accuracy approaching ceiling levels (above 97\% in most cases). Such trends affirm the model’s capacity to capture relevant features even in challenging differential diagnosis.

Conversely, classes such as Keratinization disorders, Neoplasms, others, and no diagnosis exhibit lower Top-1 performance, yet still benefit from 20–40\% absolute improvement when considering Top-5 predictions. This suggests that while these categories are harder to pinpoint as the top choice, the model often recognizes them as plausible alternatives, supporting their inclusion in a ranked differential. This reflects the inherent difficulty or class imbalance in these categories. These results highlight strong discriminative performance in clinically prevalent and visually distinct categories, while also pointing to the need for improvement in more ambiguous or underrepresented classes.

\begin{table}[h]
\scriptsize
\setlength{\tabcolsep}{3pt} % column padding
\centering
\caption{Detailed performance analysis of best models for main class (MC) prediction. All metrics include Accuracy (Top-1,3,5) and AUC (ovr) as \% for both: all classes and per class.}
\label{tab:swin_mc_results}
\begin{tabular}{c@{\hspace{3pt}}c|@{\hspace{3pt}}c|@{\hspace{3pt}}c@{\hspace{3pt}}c@{\hspace{3pt}}c@{\hspace{3pt}}c@{\hspace{3pt}}c@{\hspace{3pt}}c@{\hspace{3pt}}c@{\hspace{3pt}}c}
\toprule
\textbf{Model} & \textbf{Metric} & \textbf{All classes} & \textbf{Infectious} & \textbf{Inflammatory} & \textbf{Keratanisation} & \textbf{Neoplasms} & \textbf{No Diagnosis} & \textbf{Other} & \textbf{Pigmentary} & \textbf{Appendages} \\
\midrule
\multirow{4}{*}{\shortstack{\textbf{Swin}\\ (MC)}} 
& Top-1  & 70.88 & 77.51 & 68.51 & 14.29 & 16.67 & 0.00  & 14.29 & 73.17 & 81.19 \\
& Top-3  & 91.06 & 94.98 & 96.10 & 50.00 & 83.33 & 33.33 & 32.14 & 87.20 & 92.08 \\
& Top-5  & 94.86 & 97.13 & 98.70 & 50.00 & 83.33 & 66.67 & 60.71 & 92.07 & 97.03 \\
& AUC    & 78.36 & 84.80 & 82.76 & 66.29 & 87.05 & 55.55 & 63.57 & 91.11 & 95.79 \\
\hline

\multirow{4}{*}{\shortstack{\textbf{CBM} \\ (Concepts+\\MC)}} 
& Top-1  & 67.55 & 72.49 & 67.21 & 7.14  & 25.00 & 0.00  & 7.14  & 70.73 & 77.23 \\
& Top-3  & 91.72 & 98.33 & 96.75 & 21.43 & 41.67 & 16.67 & 32.14 & 89.02 & 90.10 \\
& Top-5  & 96.57 & 99.52 & 99.68 & 57.14 & 83.33 & 66.67 & 60.71 & 94.51 & 97.03 \\
& AUC    & 79.17 & 85.70 & 81.80 & 56.40 & 79.10 & 74.34 & 68.60 & 93.09 & 94.37 \\

\hline

\multirow{4}{*}{\shortstack{\textbf{CBM Type 2} \\(Concepts+\\MC \& SC)}} 
& Top-1  & 70.12 & 79.43 & 66.23 & 7.14  & 16.67 & 0.00  & 14.29 & 71.95 & 75.25 \\
& Top-3  & 90.49 & 94.50 & 96.43 & 14.29 & 41.67 & 33.33 & 46.43 & 87.80 & 92.08 \\
& Top-5  & 95.53 & 95.93 & 99.03 & 64.29 & 75.00 & 66.67 & 82.14 & 95.73 & 95.05 \\
& AUC    & 78.71    & 84.78 & 83.08 & 66.30 & 80.62 & 60.91 & 68.39 & 91.77 & 93.82 \\
\bottomrule
\end{tabular}
\end{table}

% The CBM Type 2 model, which jointly predicts both subclass and main class labels using a parallel concept-grounded architecture, yields the best performance across nearly all metrics. These results demonstrate the advantage of hierarchical multi-task supervision grounded in medically meaningful concepts.

% ================================
% Supplementary: Multi-disease Co-occurrence Analysis
% (Requires: \usepackage{booktabs} and optionally \usepackage{array})
% ================================

\paragraph{Multi-disease Co-occurrence Analysis.}
A small but clinically meaningful subset contains concurrent lesions from more than one disease type on the same anatomical site (e.g., \textit{Inflammatory + Fungal}, \textit{Fungal + Bacterial}). We encode these as dedicated second-level subclasses to let models explicitly learn multi-disease patterns and disentangle overlapping cues. To assess behavior on these cases, we performed a targeted misclassification analysis with the best \textit{Swin Transformer}, recording whether predictions matched both constituent types (\emph{Predicted as Both}), only one (\emph{Predicted as Either}), or neither (\emph{Predicted as Other}). Results are summarized in Table~\ref{tab:supp_multi_disease}.

\begin{table}[h!]
\scriptsize
\setlength{\tabcolsep}{3pt}
\centering
\caption{Misclassification analysis for multi-disease co-occurrence samples. ``Predicted as Both'' denotes a correct assignment to the multi-disease subclass; ``Predicted as Either'' lists counts that match one constituent type; ``Predicted as Other'' lists counts for unrelated classes.}
\label{tab:supp_multi_disease}
\vspace{4pt}
\begin{tabular}{p{3.2cm} p{2.0cm} p{4.5cm} p{2.5cm}}
\toprule
\textbf{True Class} & \textbf{Predicted as Both} & \textbf{Predicted as Either} & \textbf{Predicted as Other} \\
\midrule
Inflammatory + Infectious--Bacterial & 5   & Inflammatory (4), Bacterial (3) & Pigmentary (1) \\
Fungal + Bacterial                   & 1   & Fungal (3), Inflammatory (1)    & --- \\
Parasitic + Bacterial                & 3   & Parasitic (2), Inflammatory (1)  & --- \\
Inflammatory + Fungal                & --- & Fungal (2), Inflammatory (1)     & --- \\
\bottomrule
\end{tabular}
\end{table}

Overall, a subset of samples is correctly recognized as their multi-disease subclass (\emph{Predicted as Both}). Many are assigned to one constituent type (\emph{Predicted as Either}), likely reflecting dominance of one pathology’s visual cues (e.g., markedly scaly plaques in fungal disease). A smaller number maps to unrelated categories (\emph{Predicted as Other}). These findings underscore the dataset’s clinical realism and motivate \textbf{multi-label, context-aware} approaches to robustly handle co-occurring dermatological conditions.

\begin{figure}[!b]
  \centering
  % \vspace{-10pt}
  \includegraphics[width=1.05\linewidth]{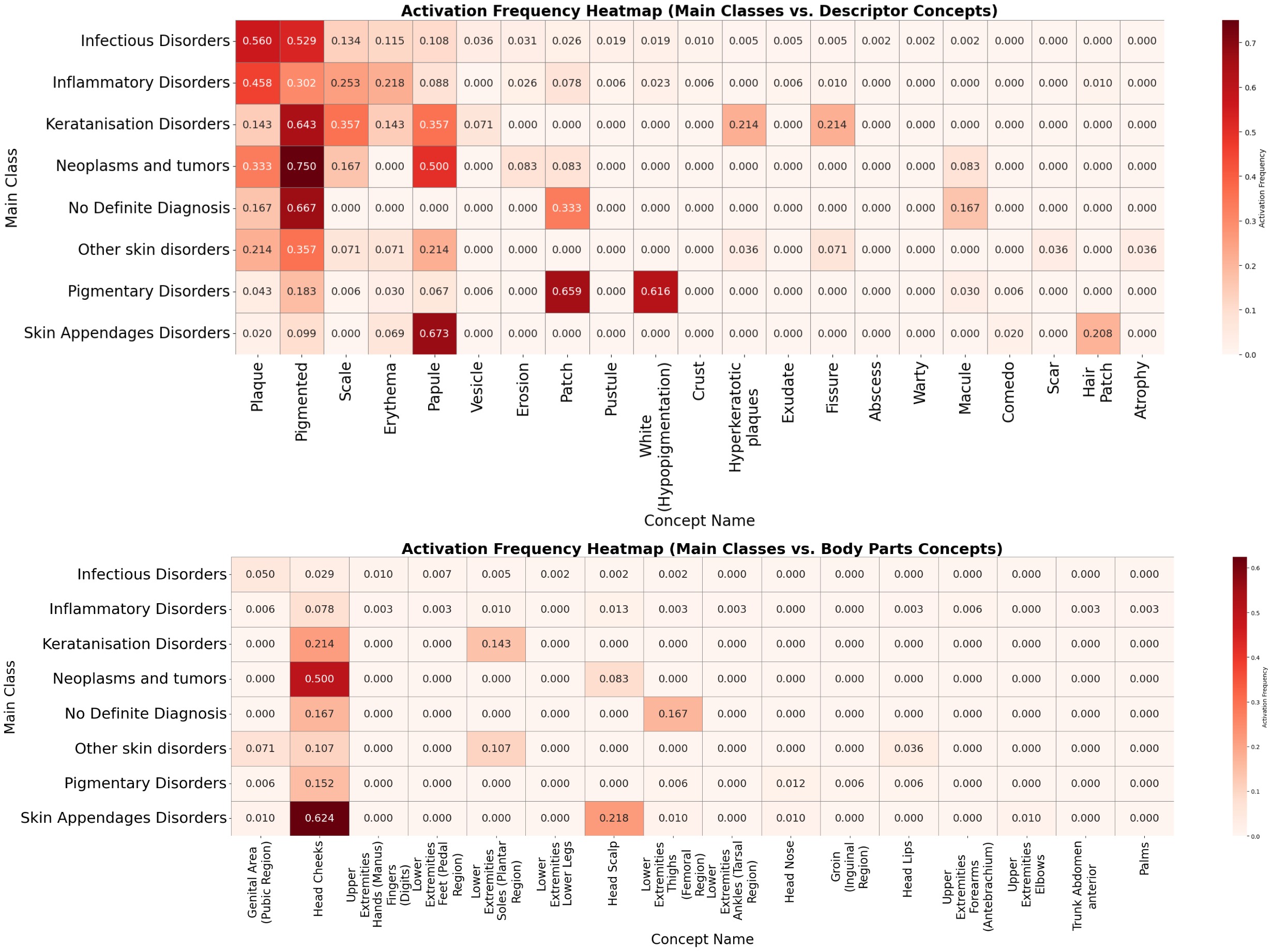}
  \vspace{-10pt}
  \caption{Concept frequency heatmaps for the analysis of active concepts per class (Main class) on the full test set.}
  % \vspace{-50pt}
  \label{fig:con_freq}
\end{figure}

\section{More examples for Post-hoc analysis}

The following post-hoc analyses are conducted using the best-performing CBM, namely the Type-2 model, on the test set provided in GitHub.

\subsection{Concept Activation Analysis}

We analyzed the activation patterns of interpretable concepts across eight dermatological disorder categories to evaluate semantic alignment in model reasoning. Concepts were grouped into two clinically motivated families: \textit{descriptor} concepts (e.g., \textit{plaque}, \textit{erythema}, \textit{vesicle}) and \textit{body part} concepts (e.g., \textit{head}, \textit{cheek}, \textit{extremities}). For each class, activation frequency was computed as the fraction of samples per class in which a given concept was predicted to be positively active and is presented as heatmaps in Sup. Fig.~\ref{fig:con_freq}. Concepts with negative activation across all classes were excluded from visualization.

Out of 47 available descriptor concepts, 21 exhibited positive activation. For body part concepts, 16 out of 49 exhibited positive activation. This indicates that the model’s decision process selectively emphasizes a small subset of clinically relevant features while disregarding many others, which could be due to lower predictive value.

Notably, descriptor activations showed meaningful alignment with known disease characteristics: \textit{papule} and \textit{pigmented} were strongly associated with \textit{Skin Appendages Disorders} and \textit{Neoplasms and Tumors} respectively, while \textit{Pigmentary Disorders} prominently activated \textit{patch} and \textit{white hypopigmentation}. In contrast, body part activations were more sparse and concentrated, with only a few regions such as \textit{head cheeks}, \textit{head scalp}, and \textit{lower extremities thighs} contributing meaningfully. The under-utilization of many anatomical concepts possibly suggests the model’s insensitivity to spatial context.

This selective concept usage underscores the need for additional constraints that promote semantic coverage and balanced concept learning. Future work could incorporate concept entropy regularization or supervision-aware attention mechanisms to encourage more uniform engagement across the concept space, particularly for underrepresented anatomical regions.

\begin{figure}
  \centering
  % \vspace{-10pt}
  \includegraphics[width=1.0\linewidth]{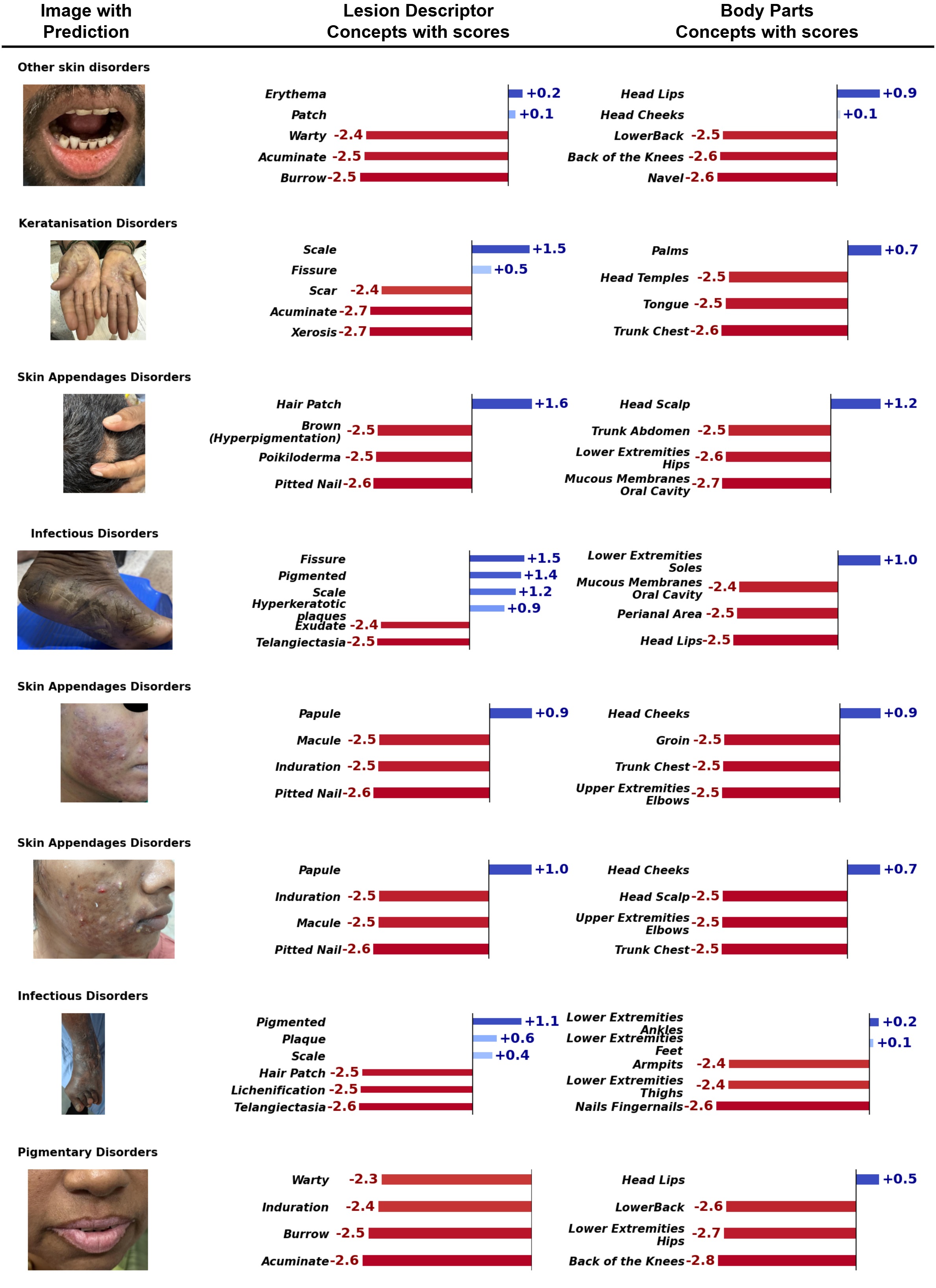}
  \vspace{-10pt}
  \caption{Bar plots showing the top and bottom-k contributing concepts (lesion descriptors and body parts) for the model’s prediction. Contribution values are computed as signed, log-scaled scores derived from the CBM’s intermediate concept logits. Blue bars indicate the concepts with positive contributions, whereas the red bars highlight the concepts with negative contributions, and all are rightly predicted. These examples specifically highlight cases where at least one body part concept has a positive contribution, with lesion descriptors also showing concurrent scores in most instances.}
  % \vspace{-50pt}
  \label{fig:ex_bp}
\end{figure}

\begin{figure}
  \centering
  % \vspace{-10pt}
  \includegraphics[width=1.0\linewidth]{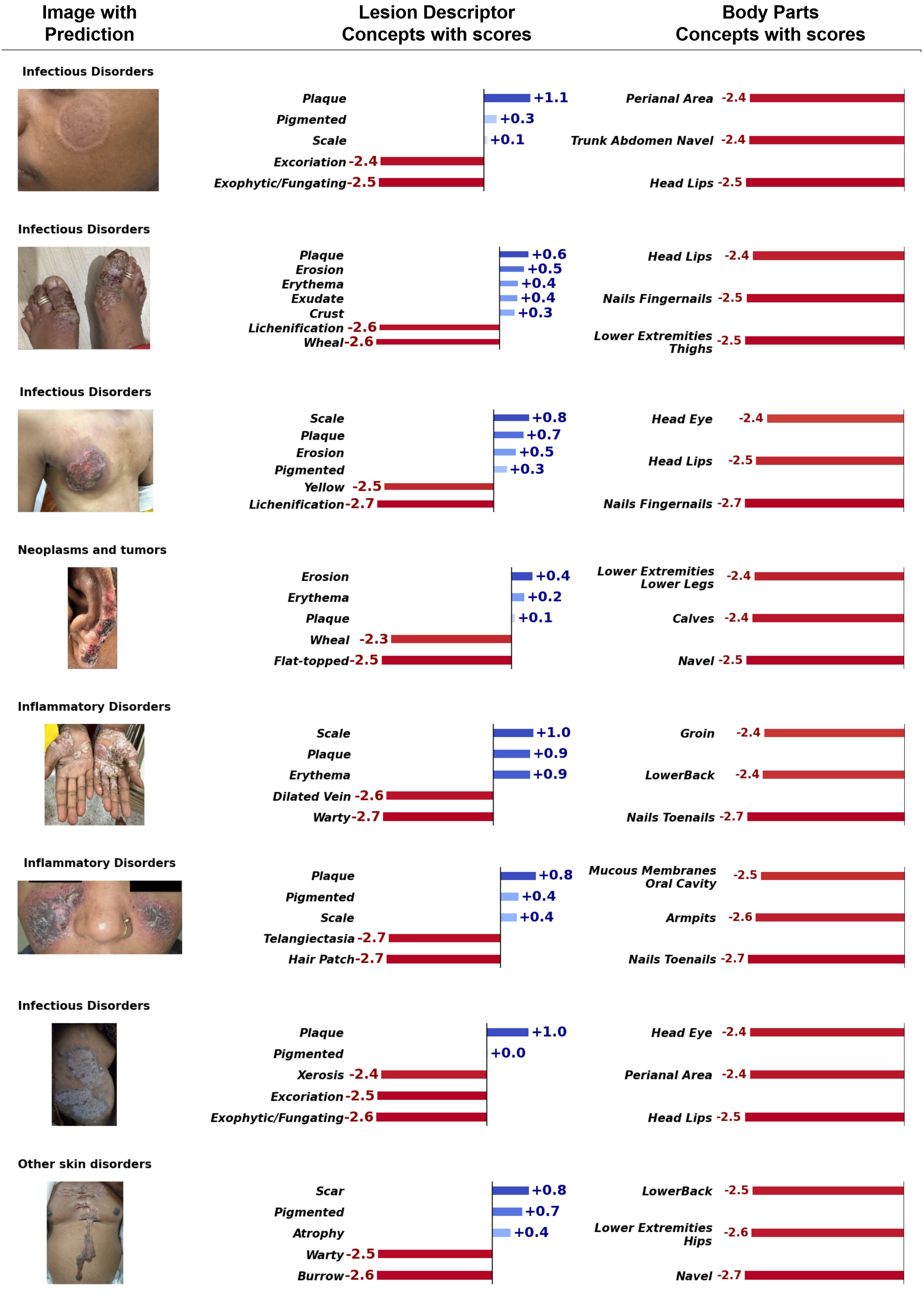}
  \vspace{-10pt}
  \caption{Bar plots showing the top and bottom-k contributing concepts (lesion descriptors and body parts) for the model’s prediction. Contribution values are computed as signed, log-scaled scores derived from the CBM’s intermediate concept logits. Blue bars indicate the concepts with positive contributions, whereas the red bars highlight the concepts with negative contributions, and all are rightly predicted. These examples specifically highlight cases where at least one lesion descriptor concept has a positive contribution, while body part concepts exhibit negative contribution.}
  % \vspace{-50pt}
  \label{fig:ex_desc}
\end{figure}

\subsection{Concept Contribution Analysis Across Semantically Disjoint Families}

To further explore the interplay between \textit{descriptor} and \textit{anatomical} concept families within our Concept Bottleneck Model (CBM), we provide two sets of illustrative examples (Sup. Fig.~\ref{fig:ex_bp}, \ref{fig:ex_desc}). These were chosen from correctly predicted samples with concept annotations verified by expert dermatologists. Contribution scores are computed using signed, log-scaled intermediate logits, reflecting each concept's influence (positive or negative) on the final class prediction.

\paragraph{(a) Positive Contribution from Body Part Concepts with Co-activation of Descriptors.}  
In Sup. Fig.~\ref{fig:ex_bp}, we show eight representative cases where at least one body part concept has a positive contribution to the model’s decision. In seven of these, we observe concurrent positive contributions from descriptor concepts, as well, suggesting that when anatomical information is utilized by the model, it is rarely used in isolation. Only one sample showed no positively contributing descriptor, which may hint at either spurious localization or weak feature learning from descriptors in that particular case. This consistent pattern of joint activation supports the view that meaningful spatial reasoning in the model emerges most reliably when supported by surface-level lesion features, reinforcing the clinical validity of concept co-activation as a desirable property in interpretable models.

\paragraph{(b) Dominance of Descriptor Concepts in Prediction with Absent Positive Anatomical Contributions.}  
Sup. Fig.~\ref{fig:ex_desc} complements the above by presenting eight cases where descriptor concepts show strong positive contributions, but none of the body part concepts make a positive impact. While both positively and negatively contributing concepts are visible, the absence of anatomical contributions even in correctly classified examples highlights the model’s stronger reliance on surface-level lesion descriptors.

These observations are consistent with the concept frequency analysis reported earlier, where descriptor concepts not only activated more frequently but also aligned more closely with disease-specific patterns. Taken together, these findings reveal a clear representational bias in the CBM toward the descriptor concept family, underscoring the need for regularization strategies or balanced supervision to ensure equitable utilization of spatial and morphological information.

% This aligns with earlier findings (main paper, Sec.~5.2) and reaffirms the model’s semantic bias.

\begin{figure}
  \centering
  % \vspace{-10pt}
  \includegraphics[width=0.8\linewidth]{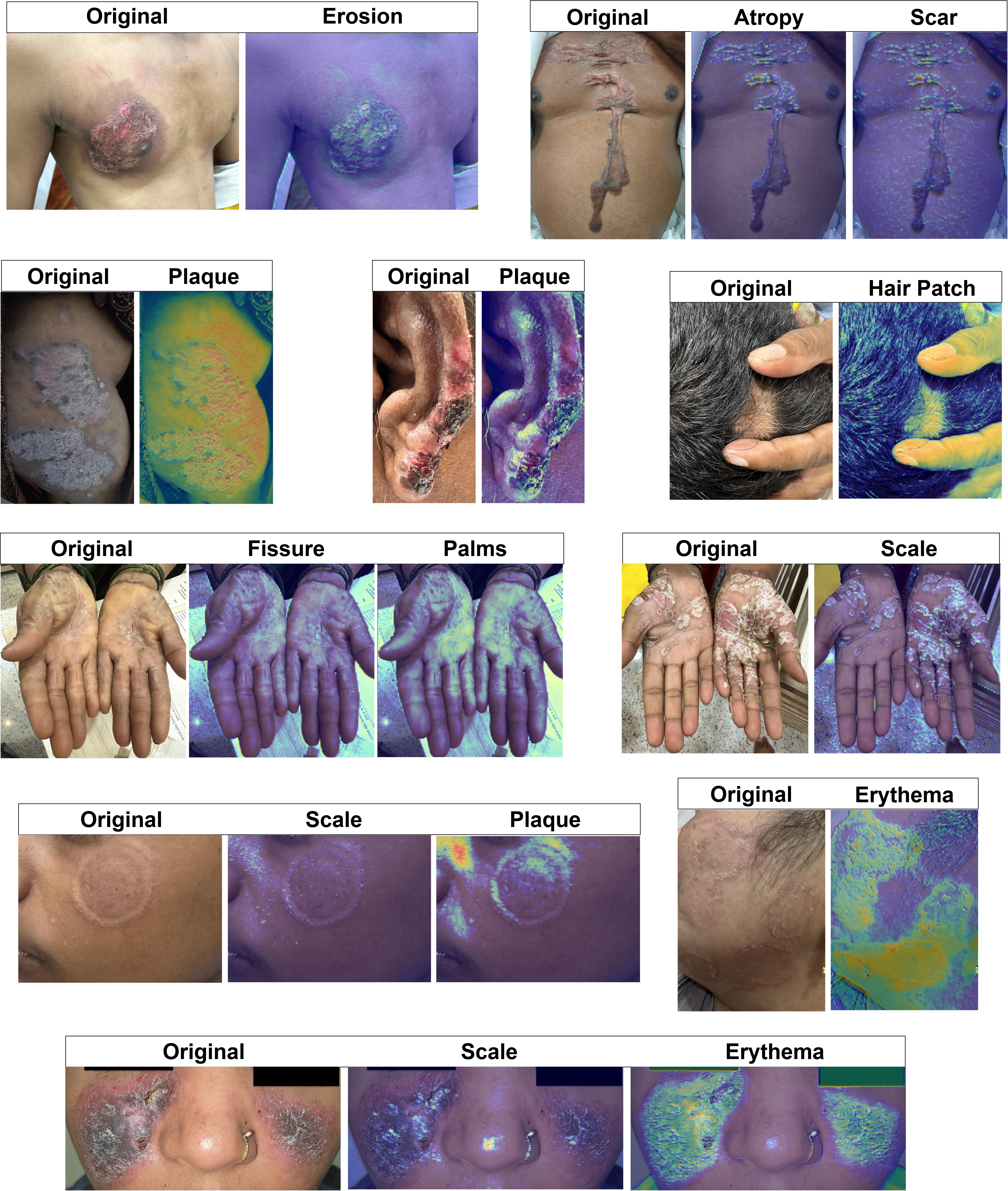}
  % \vspace{-10pt}
  \caption{Illustrations of Grad-cam visualizations over Swin Transformer by choosing a specific concept. Highlighted parts of the images with either green, yellow, or red points to the regions responsible for the positive contribution of the concept. All the images were further cross-validated by the dermatologists for the right concept activation.}
  % \vspace{-50pt}
  \label{fig:grad_more}
\end{figure}

\subsection{Concept-specific Spatial Attribution via Grad-CAM}
To gain spatial interpretability over concept activations, we performed Grad-CAM analysis on the Swin Transformer’s CBM layer for a subset of correctly classified examples. As illustrated in Sup. Fig.~\ref{fig:grad_more}, each visualization corresponds to a specific concept with positive contribution selected from the CBM’s intermediate bottleneck (e.g., \textit{plaque}, \textit{scale}, \textit{palms}), with ten examples chosen to demonstrate the range and specificity of concept-localized attention. Using logit-directed gradient backpropagation from the concept prediction head, we generated class-discriminative heatmaps that reveal the spatial regions influencing each concept’s activation. These visualizations reinforce the semantic alignment between model representations and clinical reasoning: lesion descriptors such as \textit{erythema} or \textit{plaque} consistently activate in regions of visible inflammation or raised morphology. Importantly, the ability to isolate spatial attributions per concept allows clinicians to verify not just \emph{what} the model has learned, but also \emph{where} it is looking, serving as a crucial step toward validating model trustworthiness in clinical settings.

\begin{figure}
  \centering
  % \vspace{-10pt}
  \includegraphics[width=1.0\linewidth]{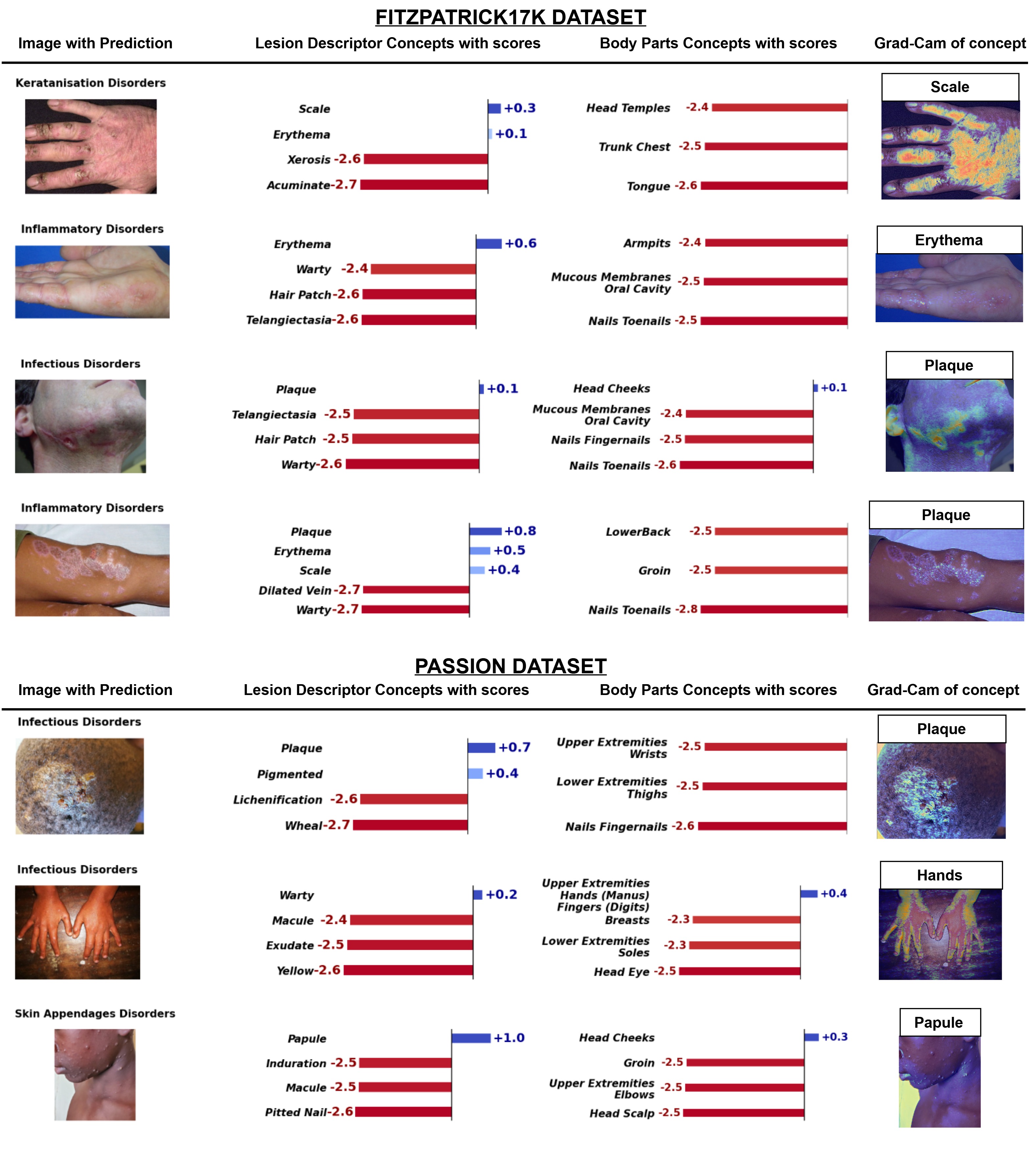}
  \vspace{-20pt}
  \caption{Illustration of interpretability analysis of  CBM (Type 2) model trained on DermaCon-IN dataset and cross validated on samples of Fitzpatrick17k and PASSION datasets. Bar plots show the top and bottom-k contributing concepts (lesion descriptors and body parts) for the model’s prediction. Contribution values are computed as signed, log-scaled scores derived from the CBM’s intermediate concept logits. Blue bars indicate the concepts with positive contributions, whereas the red bars highlight the concepts with negative contributions, and all are rightly predicted. The rightmost column presents the Grad-Cam analysis of the best concept.}
  % \vspace{-50pt}
  \label{fig:others}
\end{figure}

\section{Cross-Dataset Validation and Distributional Coverage}

Our dataset has been developed with a focus on flexibility and interoperability, aiming to support a range of downstream dermatological AI tasks. Its hierarchical taxonomy enables researchers to adjust label granularity according to their specific objectives, whether for disease classification, concept prediction, or region-aware modeling.

To demonstrate this integrative potential, we conducted cross-dataset evaluations using two publicly available benchmarks: the PASSION~\cite{gottfrois2024passion} dataset and the Fitzpatrick17k~\cite{groh2021evaluating} dataset. Using our label hierarchy, we aligned $40$ randomly selected samples (due to unavailability of val splits) from each dataset to our taxonomy, as the label space for both datasets was different.  The best performing CBM (Type 2) model, trained exclusively on DermaCon-IN dataset, correctly predicted $30$ samples from PASSION and $33$ from Fitzpatrick17k, and produced clinically valid and interpretable Grad-CAM visualizations across these external samples. All outputs were reviewed and verified by a board-certified dermatologist for both diagnostic accuracy and localisation relevance. A few samples of our analysis is presented in Sup. Fig.~\ref{fig:others}.

These observations suggest that our dataset is neither isolated nor out-of-distribution. Rather, it addresses a critical representational gap by contributing cases from underrepresented skin tones, outpatient clinical settings, and real-world diagnostic variation specific to South Asian populations.

% While further large-scale benchmarking is necessary, these results indicate that the structure and diversity of our dataset offer practical utility for researchers working with heterogeneous dermatological datasets.
% Though we do not claim superior generalisation, our early validation suggests that models trained on our dataset may exhibit improved transferability across datasets and patient populations.
While further large-scale benchmarking is necessary, our results suggest that the structure and diversity of the dataset provide practical value for researchers dealing with heterogeneous dermatological data. Although we do not claim generalisability, the dataset can be an asset for training models aimed towards transferability and foundational learning.

\section{Ethical Considerations}
% Discuss any known limitations, biases, or ethical aspects relevant to your study.

This dataset was curated through clinical data collection from consenting patients from outpatient clinics. Ethical diligence was integrated into the dataset lifecycle, including data collection, annotation, privacy protection, and intended use, in line with emerging best practices for responsible dataset curation in human-centric computer vision (HCCV). The protocol was reviewed and approved by the institutional ethics committee.

\paragraph{Data Source and Informed Consent}
All images were collected during routine dermatological consultations with informed consent from patients in the native language or a language more easily understood by the patients. No data was scraped from the web or obtained from public platforms.  

\paragraph{Privacy Protection and Anonymization}
We ensured that no personally identifiable information (PII) was present in any image and is irreversibly coded for identity. Facial identifiable features such as eyes, tattoos, and other identifiable marks were excluded or cropped from the images. Additionally, all embedded metadata (e.g., timestamps, device IDs, location data) was removed. Participants were informed of their right to withdraw from the study.

% \paragraph{Representation and Fairness}
% To address demographic gaps in dermatology datasets, we focused on capturing skin tones, disorders, and clinical presentations prevalent in the Indian subcontinent. Every image was annotated with Fitzpatrick Skin Type (III–VI) and Monk Skin Tone (MST) labels to facilitate fairness audits across tonal gradients. We retained epidemiologically realistic class distributions.
\paragraph{Cultural Sensitivity.}
During image collection involving female patients, all study procedures were explained by, or in the presence of, female medical staff to ensure comfort, privacy, and culturally appropriate engagement.

\paragraph{Annotation Integrity and Labeling Ethics}
 All annotations were carried out by trained clinical annotators working under a dermatologist's supervision. We deliberately avoided inferring any identity-related attributes from images. Skin tone annotations were made directly based on non-affected skin regions using standardized MST/Fitzpatrick reference scales. Inter-annotator consistency was maintained through calibration sessions, and difficult cases were adjudicated by a senior dermatologist.

\paragraph{Use Restrictions and Responsible Deployment}
The dataset is intended for academic research, particularly in studying fairness, robustness, and interpretability in dermatological disease classification. It is not intended for commercial use, biometric profiling, identity inference, or deployment in real-time diagnostic tools without regulatory and clinical validation. We intend to make the dataset openly accessible to support research in dermatological AI. However, access will be granted following a background verification process to ensure responsible and ethical use.

\section{Societal Impact}
The potential impact of this dataset lies not in the scale of data amassed, but in its grounding within everyday clinical reality. By capturing routine skin conditions in South Asian outpatient settings, where diagnostic decisions often occur with limited advanced imaging, the dataset reflects how dermatological care is actually practiced in many parts of the world. Its value, therefore, is not just in building algorithms, but in offering a testbed where machine learning systems can be evaluated under conditions that mirror frontline clinical environments. Additionally, by organizing information through interpretable clinical descriptors and anatomical context, the dataset invites collaboration between clinicians and model developers, not as a retrospective audit, but as an integral part of design. This is particularly meaningful for low-resource settings where human expertise must be complemented. Rather than aiming for universal solutions, we hope this dataset contributes to a shift toward more situated, dialogic approaches to clinical AI development.

Beyond its immediate utility for algorithm training, the dataset also raises possibilities for advancing how uncertainty and variability are treated in clinical AI. By including lesions with atypical or overlapping features, cases that might otherwise be excluded, it encourages models to engage with diagnostic edge cases rather than avoid them. This is particularly relevant in primary care and peripheral settings, where clear-cut presentations are the exception, not the norm. Additionally, the dataset provides opportunities to study how different combinations of features influence clinical suspicion, enabling future work on model calibration and risk stratification. In this way, the dataset serves not only as input to predictive systems and as a scaffold for developing tools that assist with triage, escalation decisions, or patient education — areas where uncertainty is not a flaw but a central part of the task.

% \section{Additional Figures and Tables}

\section{Glossary of Terms}

\begin{longtable}{>{\raggedright\arraybackslash}p{2.5cm} >{\raggedright\arraybackslash}p{11cm}}
\caption*{Glossary continued across pages.}\\
\toprule
\textbf{Term} & \textbf{Definition} \\
\midrule
\endfirsthead

\multicolumn{2}{l}{\textit{Continued from previous page}} \\
\toprule
\textbf{Term} & \textbf{Definition} \\
\midrule
\endhead

\midrule \multicolumn{2}{r}{\textit{Continued on next page}} \\
\endfoot

\bottomrule
\endlastfoot

\textbf{Alterations} & Alterations denote clinically observable changes in lesion attributes such as color, texture, or size over time, often indicating progression, regression, or therapeutic response. \\

\textbf{Anatomical} & Anatomical refers to the specific bodily region where a lesion appears, which can influence diagnostic reasoning due to region-specific disease prevalence and morphology. \\

\textbf{Anomalies} & Anomalies represent structural or morphological deviations from typical presentation, often indicating pathological states like vascular malformations or congenital disabilities. \\

\textbf{Appendageal} & Appendageal pertains to skin-associated structures such as hair follicles, sebaceous glands, and nails, diagnostically relevant in conditions like alopecia or onychomycosis. \\

\textbf{Atopic} & Atopic describes conditions driven by hypersensitivity or allergic predisposition, notably atopic dermatitis, often linked to genetic and environmental triggers. \\

\textbf{Blistering} & Blistering refers to fluid-filled skin elevations (vesicles/bullae). \\

\textbf{Chronic} & Chronic refers to a Long-standing or recurrent condition; affects annotation and model expectations. \\

\textbf{Corticosteroid} & Corticosteroid refers to the Anti-inflammatory drug class; often implicated in misuse or treatment. \\

\textbf{Curricula} & Curricula refer to Formal educational content; they provide background for how clinicians interpret and label data. \\

\textbf{Cutaneous} & Relating to the skin; distinguishes from mucosal involvement. \\

\textbf{Demographic} & Relating to population traits; crucial for fairness and representation. \\

\textbf{Dermatological} & About the skin and its diseases; defines the scope of the dataset. \\

\textbf{Dermatophytosis} & Dermatophytosis refers to a Fungal infection of skin, hair, or nails. \\

\textbf{Dermatoses} & Dermatoses refers to a general term for skin diseases. \\

\textbf{Dermis} & Dermis refers to the middle layer of skin, the site of many dermatological processes. \\

\textbf{Dermoscopy} & Dermoscopy is a non-invasive skin imaging technique for assessing pigmented lesions. \\

\textbf{Diagnostic} & Related to identifying diseases; informs how images are labeled. \\

\textbf{Endemic} & Regularly found among a population; informs dataset collection region. \\

\textbf{Epidemiologically} & In terms of disease patterns in populations; informs dataset design. \\

\textbf{Epidermis} & Epidermis refers to the outer skin layer, involved in superficial lesions and visual markers. \\

\textbf{Erosion} & Erosion refers to the loss of part of the epidermis; visible and diagnosable via image. \\

\textbf{Etiology} & The cause or origin of a disease; critical for understanding disease mechanisms in datasets. \\

\textbf{Exogenous} & Exogenous refers to originating outside the body; affects classification of environmental skin damage. \\

\textbf{Fluctuant} & Fluctuant refers to Soft and compressible; indicates the presence of fluid (e.g., cysts, abscesses). \\

\textbf{Granulomatous} & Related to granuloma formation; chronic inflammation marker. \\

\textbf{Hematological} & Related to blood or blood-forming organs; systemic diseases may manifest cutaneously. \\

\textbf{ICD} & ICD refers to the International Classification of Diseases; a standardized coding system used globally for diagnosis and reporting. \\

\textbf{Incontinence} & In dermatology, refers to pigment incontinence where melanin leaks into the dermis. \\

\textbf{Infectious} & Caused by pathogens; represents a major disease category in dermatology. \\

\textbf{Inflammatory} & Involving immune response; another common skin disease category. \\

\textbf{Keratinization} & Keratinization refers to skin thickening; central in conditions like psoriasis. \\

\textbf{LMIC} & LMIC refers to Low- and Middle-Income Countries; refers to geographic and economic contexts. \\

\textbf{Lesion} & Lesion refers to any abnormal area on the skin, the primary subject of dermatological datasets. \\

\textbf{Melanin} & Melanin refers to a pigment responsible for skin color; key in diagnosing pigmentation disorders. \\

\textbf{Melanoma} & Melanoma refers to a malignant tumor of melanocytes, a key example of a neoplastic skin lesion. \\

\textbf{Morbidity} & Morbidity refers to the rate of disease in a population; it helps contextualize dataset relevance. \\

\textbf{Morphological} & Morphological refers to Concerned with form and structure; describes visual features of lesions. \\

\textbf{Mucosal} & Mucosal refers to moist linings (e.g., inside mouth); important in systemic diseases. \\

\textbf{Neoplastic} & Neoplastic refers to abnormal cell growth; includes benign and malignant tumors. \\

\textbf{Ontologies} & Ontologies refer to Structured vocabularies linking concepts; useful for clinical data labeling. \\

\textbf{Outpatient} & Outpatient refers to a Healthcare setting where patients are not admitted overnight; a common source of dermatology images. \\

\textbf{Palpable} & Able to be felt; clinical term for raised or solid lesions. \\

\textbf{Pathological} & Related to disease processes; central to diagnosis and data labeling. \\

\textbf{Pathophysiology} & Pathophysiology refers to functional changes associated with disease; it links imaging to mechanisms. \\

\textbf{Phenotypic} & Phenotypic refers to observable traits or characteristics; crucial for image-based annotations. \\

\textbf{Pigment} & Pigment refers to coloring matter in skin; changes indicate various disorders such as vitiligo or melasma. \\

\textbf{Pigmentary} & Pigmentary refers to relating to skin color changes; includes hyper- and hypo-pigmentation. \\

\textbf{Pigmentary Alterations} & Pigmentary Alterations refers to Changes in skin color, important for diagnosing pigmentation disorders. \\

\textbf{Polymorphic} & Polymorphic refers to having varied forms; describes diverse visual patterns of lesions. \\

\textbf{Primary Lesions} & Primary Lesions refers to Initial skin changes (e.g., macule, papule); basis for clinical diagnosis. \\

\textbf{Pruritic} & Pruritic refers to Itchy; common symptom that guides diagnosis. \\

\textbf{Rook's} & Rook's refers to Rook’s Textbook of Dermatology; a key reference in dermatological classification and clinical teaching. \\

\textbf{Secondary Changes} & Secondary Changes refer to lesion alterations due to disease progression or external factors. \\

\textbf{Subcutaneous} & Subcutaneous refers to the layer beneath the dermis composed of fat and connective tissue; affected in deep infections or nodules. \\

\textbf{Systemic} & Systemic refers to affecting the entire body; differentiates skin manifestations of internal diseases. \\

\textbf{Taxonomies} & Taxonomies refer to systematic classification of concepts or conditions; essential for dataset organization and ontology mapping. \\

\textbf{Varicosities} & Varicosities refer to dilated veins; visible skin features relevant in elderly or vascular conditions. \\

\textbf{Vascular} & Vascular refers to relating to blood vessels; includes lesions like purpura and telangiectasia. \\

\textbf{Vasculitis} & Vasculitis refers to the Inflammation of blood vessels; it can present with palpable purpura. \\

\textbf{Vesiculobullous} & Vesiculobullous refers to Diseases characterized by vesicles and bullae (e.g., pemphigus). \\

\textbf{Xanthomas} & Xanthomas refers to Lipid-rich lesions; indicative of metabolic disorders. \\

\textbf{Xerotic} & Xerotic refers to Dry skin; common in aging and environmental dermatitis. \\

\end{longtable}

% \newpage

% \input{main.bbl}    

% \end{document}
% \newpage

% {\small
% \bibliography{ref}  % do not include .bib extension
% }

\end{document}